\documentclass[pra,twocolumn,tightenlines,showpacs,nofootinbib,superscriptaddress]{revtex4-1}
\usepackage{color}
\usepackage{bm,dcolumn,amsmath,graphicx,amsfonts,amssymb,multirow}

\newcommand{\JRME}[1]{\langle J\!\parallel\!#1\!\parallel\! J' \rangle}

\newcommand{\FRME}[1]{\langle F\!\parallel\!#1\!\parallel\! F' \rangle}
\newcommand{\mean}[1]{\langle #1\rangle}
\newcommand{\eo}{\varepsilon_{\rm o}}
\newcommand{\mo}{\mu_{\rm o}}
\newcommand{\w}{\omega}
\newcommand{\wo}{\omega_{\rm o}}
\newcommand{\G}{\Gamma}
\renewcommand{\L}{\mathcal{L}}
\newcommand{\V}{\mathcal{V}}

\newcommand{\R}{\mathcal{R}}
\newcommand{\Dop}{\Delta\omega_{\rm D}}
\newcommand{\phipnc}{\varphi_{_{\mathrm{PNC}}}}

\newcommand{\sixj}[6]{ \begin{Bmatrix}
  #1 & #2 & #3 \\
  #4 & #5 & #6
 \end{Bmatrix}}

\begin{document}
\title{Fundamentals of Cavity-Enhanced Polarimetry for Parity-Nonconserving Optical Rotation Measurements: Application to Xe, Hg and I.}
\author{L. Bougas}
\affiliation{Institute of Electronic Structure and Laser, Foundation for Research and Technology-Hellas 71110 Heraklion-Crete, Greece}
\affiliation{Department of Physics, University of Crete 71003 Heraklion-Crete, Greece}
\author{G. E. Katsoprinakis}
\affiliation{Institute of Electronic Structure and Laser, Foundation for Research and Technology-Hellas 71110 Heraklion-Crete, Greece}
\author{W. von Klitzing}
\affiliation{Institute of Electronic Structure and Laser, Foundation for Research and Technology-Hellas 71110 Heraklion-Crete, Greece}
\author{T. P. Rakitzis}
\email{ptr@iesl.forth.gr}
\affiliation{Institute of Electronic Structure and Laser, Foundation for Research and Technology-Hellas 71110 Heraklion-Crete, Greece}
\affiliation{Department of Physics, University of Crete 71003 Heraklion-Crete, Greece}
\date{\today}

\begin{abstract}
We present the theoretical basis of a cavity-enhanced polarimetric scheme for the measurement of parity-nonconserving (PNC) optical rotation. We discuss the possibility of detecting PNC optical rotation in accessible transitions in metastable Xe and Hg, and ground state I. In particular, the physics of the PNC optical rotation is presented, and we explore the lineshape effects on the expected PNC optical rotation signals. Furthermore, we present an analysis of the eigenpolarizations of the cavity-enhanced polarimeter, which is necessary for understanding the measurement procedure and the ability of employing robust background subtraction procedures using two novel signal reversals. Using recent atomic structure theoretical calculations, we present simulations of the PNC optical rotation signals for all proposed transitions, assuming a range of experimentally feasible parameters. Finally, the possibility of performing sensitive measurements of the nuclear-spin-dependent PNC effects is investigated, for the odd-neutron nuclei $^{129}$Xe and $^{199}$Hg, and the odd-proton nucleus $^{127}$I.
 \end{abstract}

\maketitle
\graphicspath{{Figs/}}
\section{Introduction}\label{sec:I}
The possibility of measuring parity non-conservation (PNC) in atoms was first considered by Zeldovich in 1959\,\cite{Zeldovich}. However, atomic PNC experiments begun only after Bouchiat and Bouchiat showed that parity mixing of atomic states scales as $\sim\rm{Z}^3$ and measurable signals could be obtained for high-Z atoms\,\cite{Bouchiat1}. For high-$Z$, the degree of $s$-$p$ parity mixing in some atomic states, is of order $10^{-12}-10^{-10}$. The precise measurement of this atomic PNC can provide a stringent low-energy test of the standard model\,\cite{Diener}, of inter-nucleon weak interactions, and of nuclear structure\,\cite{Ginges}. \\
\indent There are several methods in which the parity mixing can be measured, and for each method the optimal atomic candidates are usually different. For example, the Stark interference technique has been used to measure PNC in Cs\,\cite{Wieman}, Yb\,\cite{Tsigutkin}, and Dy\,\cite{Nguyen}, and proposed for Fr\,\cite{Bouchiat2008} and Rb\,\cite{Sheng2010,DzubaRb2012}; the optical rotation technique has been used successfully for Tl\,\cite{Vetter}, Bi\,\cite{McPherson}, and Pb\,\cite{Meekhof1}; the ac-Stark shift method is proposed for Ba$^+$\,\cite{Sherman05} and Ra$^+$\,\cite{Wansbeek,Nunez2013} ions; and the hyperfine transition method is proposed for K\,\cite{Potassium}, Rb\,\cite{Sheng2010}, and Fr\,\cite{Gomez2006}. To date, the most successful atomic PNC measurement has been the 0.35\% precision measurement of nuclear-spin-independent PNC in Cs\,\cite{Wieman}. As the precision in the atomic theory of other PNC candidates is not expected to significantly surpass the theoretical precision of Cs, current experiments are aiming at other important goals that are not dependent on extremely precise atomic theory calculations. Examples include the measurement of atomic PNC on a chain of isotopes\,\cite{Dzuba1986,Fortson1990}, and the measurement of nuclear spin-dependent effects\,\cite{Ginges}. Therefore, along these lines, PNC experiments are in progress as mentioned above\,\cite{Tsigutkin, Nguyen, Sherman05, Nunez2013}.\\
\indent Due to the difficulty of controlling all the relevant parameters to the required precision, there have been only a handful of successful atomic PNC measurements, and even the few successful experiments have typically required 10-20 years to yield precise results \,\cite{Wieman, Vetter, Tsigutkin}. In addition, some of the current atomic PNC experiments are no longer tabletop, as they are performed on radioactive isotopes with short half lives of a few minutes, such as on Fr at TRIUMF\,\cite{Gomez2006} and Ra$^+$ at KVI \,\cite{Wansbeek,Nunez2013}.\\
\indent Recently, our group has proposed an extension of the optical rotation technique, with the use of a novel bow-tie optical cavity\,\cite{Bougas}. We show in detail that the proposed cavity-enhanced technique produces large experimental optical rotation signals and robust experimental checks, and allows new atomic candidates to be considered. Specifically, our proposal has several potential advantages, which solve some of the problems of past PNC optical rotation experiments. These advantages include the following:\\
\indent (a) The effective optical-rotation pathlength is enhanced using a high-finesse cavity, by $2\mathcal{F}/\pi$ where $\mathcal{F}$ is the finesse of the cavity (for high-finesse cavities, $\mathcal{F}\sim10^4-10^5$), allowing the study of PNC in atomic systems for which single-pass optical rotation from available column densities is otherwise too small. We focus on metastable states in Hg and Xe\,\cite{Bougas}, and ground-state\,I atoms\,\cite{PNCI}, for which the single-pass optical rotation from available column densities require enhancement of between $10^2$-$10^4$ cavity passes to produce measurable signals. In addition, the proposed atomic systems are compatible with a high-finesse optical cavity, as high atomic densities can be produced at around room temperature (for the case of Tl, Bi, and Pb, temperatures in excess of 1000\,K were required, which is difficult to combine with high-transmission windows and a stable optical cavity).\\
\indent (b) Two novel signal reversals are introduced. The main limitation in the original optical rotation experiments was the lack of rapid subtraction procedures or signal reversals. The proposed signal reversals are effected either by inverting the longitudinal magnetic field in the cavity, or by shifting the cavity resonance to an opposite polarization mode. These signal reversals can be performed at a high repetition rate, and allow the absolute optical rotation to be measured, without needing to remove the gas sample from the cavity. In addition, as metastable Hg and Xe and ground-state I can be produced by optical pumping, photodissociation or electrical discharge, the concentration of these species can be varied very quickly, giving an additional rapid subtraction procedure.\\
\indent (c) Of the proposed PNC candidates, both Hg and Xe have several, commercially available, stable isotopes. In addition, Hg and Xe each have two isotopes with an odd-neutron nucleus ($^{199}$Hg, $^{201}$Hg, and $^{129}$Xe, $^{131}$Xe). Moreover, iodine has a radioactive isotope, $^{129}$I, which can be commercially obtained. Both I isotopes have an odd-proton nucleus. Therefore, nuclear spin-dependent effects can be measured for both odd-neutron and odd-proton nuclei, as well as PNC measurements along a chain of isotopes\,\cite{Dzuba1986, Fortson1990, Brown2009}. \\
\indent The aim of this paper is to explain the main features of the cavity-enhanced PNC optical rotation scheme in depth and to present simulated experimental signals for Hg, Xe, and I. In Section\,\ref{sec:II} we describe in brief the origin of the PNC optical rotation. In Section\,\ref{sec:AtomicSystems} we introduce the atomic systems considered for future PNC investigations using the optical rotation technique. In addition, we examine the experimental feasibility of PNC measurements in these atomic systems. In Section\,\ref{sec:III} we describe the properties of the cavity-enhanced scheme and derive the eigenmodes of a bow-tie cavity with circular birefringence (Faraday rotation and PNC optical rotation) and linear birefringence, and discuss how the signal reversals are implemented. In Section\,\ref{sec:IV} we simulate the PNC lineshapes for several transitions in Hg, Xe, and I, for a range of experimental conditions, and discuss the results.\\

\section{PNC Optical Rotation}\label{sec:II}
\indent In this section we present the physics of the PNC optical rotation technique. We note that the equations appearing here are expressed in S.I. units and the presented formulas follow largely the structure of Ref.\,\cite{Sand,Vet} with helpful material coming from Refs.\,\cite{Sob,Ver,Rus,Sid,Pur,Kat,Steck}. In addition, the derivation of the PNC rotation angle also draws from Refs.\,\cite{Ginges,DzuFlaXeHg,ForLew}.\\
\newline
\indent A PNC neutral-current interaction between the electrons and the nucleus of an atom, mixes the parity eigenstates of the atom. This PNC-induced mixing allows for a weak electric-dipole transition, with amplitude $E1_{\rm{PNC}}$, between states of the same parity. The size of $E1_{\rm{PNC}}$ increases approximately as $\sim\rm{Z}^3$ and is inversely proportional to the energy difference between the states of opposite parity mixed by the weak interaction\,\cite{Bouchiat1}, and typically is of order 10$^{-11}$-10$^{-10}\,\rm{e}\alpha_B$ ($e$ is the charge of the electron, and $\alpha_B$ the Bohr radius). Measurement of this small parity nonconserving amplitude is achieved through its interference with a larger parity conserving amplitude. \\
\indent In a PNC optical rotation experiment, the parity conserving amplitude is an allowed magnetic-dipole amplitude $M1$. The interference between the dominant $M1$ allowed amplitude and the PNC-induced $E1_{\rm{PNC}}$ amplitude leads to optical activity. The PNC-induced optical rotation $\phipnc$ arises due to the difference in the indices of refraction for left- and right-circular polarized light in the vicinity of the magnetic dipole resonance:
\begin{equation}\label{eq:phinq}
\phipnc = \frac{\w l}{c} \frac{n'_+ - n'_-}{2} = \frac{\pi l}{\lambda} (n'_+ - n'_-),
\end{equation}
where $l$ is the length of vapor, $\lambda$ is the optical wavelength, $\omega$ is the optical frequency, and $n^{\prime}_{\pm}$ are the real parts of the refractive indices for left- and right-circular polarized light respectively (which are functions of the optical frequency $\omega$). \\
\subsection{$M1$ Magnetic dipole interaction}
\indent We assume a magnetic dipole interaction of a laser beam with an atomic vapor. Treating the transition as a damped oscillator with a damping factor $\G$, the index of refraction can be put in the form:
\begin{equation}\label{eq:ngen}
n  = n' + i n'' = 1 + \frac{\pi\mo e^2}{4 m \wo}\: \rho f\: \L (\w-\wo),
\end{equation}
\noindent where $\wo$ is the resonant transition frequency, $\rho$ the vapor density, $f$ the oscillator strength and $\L=\L'+i\L''$ the Lorentz lineshape function (given in Eqs. \eqref{eq:lorentzD} and \eqref{eq:lorentzA}). Assuming that the transition is an isolated $J\rightarrow J'$ line without hyperfine structure, we have:
\begin{equation}\label{eq:f}
f= \frac{2\, m\, \w_{\rm o}}{3\, \hbar\, e^2}\: \frac{M1^2}{2 J+1},
\end{equation}
\noindent where $M1\equiv\mean{M1} \equiv \JRME{\mu^{(1)}}$ is the reduced matrix element for the magnetic dipole operator. Taking into account the Doppler broadening mechanism of the thermal vapor (see Appendix \ref{app:LorDop}), and using Eq.\,\eqref{eq:f}, we can put Eq.\,\eqref{eq:ngen} in the form:
\begin{equation}\label{eq:nmatel}
n = 1 + \frac{\pi\mo}{2 \hbar}\: \frac{\rho}{2 J+1}\: \frac{M1^2}{3}\: \V(\w-\wo).
\end{equation}
The Voigt profile functions in $\V = \V' + i \V''$ are given in Eqs. \,\eqref{eq:voigtA} and \eqref{eq:voigtD}. Assuming a non-zero nuclear spin, $I$, we must take into account the hyperfine structure. Using:

\begin{widetext}
	\begin{equation}\label{eq:FvsJRME}
		\FRME{T^{(k)}}=(-1)^{I+k+J+F'}\sqrt{\left(2 F+1\right)\left(2 F'+1\right)}\sixj{J}{k}{J'}{F'}{I}{F} \JRME{T^{(k)}},
	\end{equation}
\end{widetext}

\noindent where $k$ is the tensor rank of the operator $T$, and from the fact that the population density of the ground hyperfine state $F$ is:

\begin{equation}
\rho(F)=\frac{2 F+1}{(2 J+1)(2 I+1)}\, \rho,
\end{equation}

\noindent then, from Eq. \eqref{eq:nmatel}, we get by summing over final states and averaging over initial states:

\begin{equation}\label{eq:nF}
n = 1 + n_{\rm o}\:\sum_{F,F'} C_{FF'}\: \V_{FF'}(\w),
\end{equation}

\noindent where we have defined:

\begin{align}\label{eq:no}
n_{\rm o}&= \frac{\pi\,\mo}{2\, \hbar}\: \frac{\rho}{2 J+1}\: \frac{M1^2}{3},\\
\label{eq:CFF}
C_{FF'}  &= \frac{\left(2 F+1\right)\left(2 F'+1\right)}{2 I+1}\sixj{J}{1}{J'}{F'}{I}{F}^2, 
\end{align}
and $\V_{FF'}(\w)\equiv\V(\w-\w_{FF'})$ for a specific $F\rightarrow F^{\prime}$ transition. Note that $n_{\rm o}$ is not a dimensionless quantity.\\
\subsection{$E1_{\rm PNC}$ electric dipole interaction}
\indent The PNC-induced electric dipole term is included in the above formulas by performing the following substitution in Eq.\,\eqref{eq:nF} and Eq.\,\eqref{eq:no}:

\begin{equation}
\frac{M1^2}{3}\rightarrow \mid\!\JRME{q\, i\, d^{(1)}_q+\mu^{(1)}_q}\!\mid^2,
\end{equation}

\noindent where $d^{(1)}$ is the electric dipole operator, $q=\pm1$, and the $i$ ensures that $E1_{\rm PNC}\equiv\mean{E1_{\rm PNC}}\equiv\JRME{id^{(1)}}$ is purely imaginary\cite{Bouchiat1}. \\
\indent The difference between the refractive indices for left- ($\sigma^+$) and right ($\sigma^-$) circularly polarized light is proportional to:

\begin{equation}\label{eq:dnPNC}
n_+ - n_- \propto 2\,i\, M1 \left(E1_{\rm PNC}-E1_{\rm PNC}^*\right) = -4 M1^2 \R,
\end{equation}

\noindent where we used $E1_{\rm PNC}^*=-E1_{\rm PNC}$ and introduced the factor $\R$:

\begin{equation}\label{eq:R}
\R\equiv{\rm Im}\left(\frac{E1_{\rm PNC}}{M1}\right).
\end{equation}

Using Eq.\,\eqref{eq:phinq} and Eq.\,\eqref{eq:dnPNC},  the PNC optical rotation angle is given by:

\begin{equation}\label{eq:phiPNC}
\phipnc  = -\frac{4 \pi l}{\lambda} \:[n(\omega)- 1]\:\R
\end{equation}

\noindent where $n(\omega)$ is the index of refraction of the medium (Eq. \eqref{eq:nF}) which is a function of the transition frequency $\omega$. The proportionality relation between the PNC optical rotation angle $\phipnc$ and the ratio $\R$ serves as the basis for this experimental technique.\\
\newline
\indent Note that the corresponding electric dipole formulas for Eqs.\,\eqref{eq:f},\eqref{eq:ngen} and \eqref{eq:no}, are obtained simply by substituting $\mo \rightarrow 1/\eo$ and $\mean{M1}\rightarrow\mean{E1}$ (with $\mean{M1}$ in $\mu_{\rm B}$ and $\mean{E1}$ in $e \alpha_{\rm o}$).

\subsection{Nuclear spin-dependent PNC effects - Anapole moment}
\indent Nuclear spin-dependent (NSD) contributions to the atomic parity violation arise due to: (a) neutral weak-current interactions between the electron and the nucleus\,\cite{Novikov}, (b) electromagnetic interaction of the electron with the nuclear anapole moment\,\cite{Khriplovich1980}, and (c) spin-independent electron-nucleon weak interactions combined with magnetic hyperfine interactions\,\cite{Flambaum1985k2}. These contributions can be included in a dimensionless constant $\varkappa$, proportional to the strength of the NSD-PNC interaction\,\cite{Ginges,Flambaum1985,AnapoleFlam1984}: 
\begin{equation}\label{eq:varkappa}
\varkappa=\varkappa_A-\frac{\mathcal{K}-1/2}{\mathcal{K}}\varkappa_2+\frac{I+1}{\mathcal{K}}\varkappa_{\mathcal{Q}_{\rm W}},
\end{equation}
where $\mathcal{K}=(-1)^{I+\frac{1}{2}-l}(I+1/2)$ ($l$ is the orbital angular momentum of the valence nucleon), $\varkappa_2\approx-0.05$\,\cite{Ginges,Flambaum1985k2} corresponds to the weak neutral currents, $|\varkappa_{\mathcal{Q}_{\rm W}}|\approx0.02$\,\cite{Ginges} appears as a radiative correction to the NSI part, and $\varkappa_A$ is the nuclear anapole moment contribution to the NSD-PNC effects.\\
\indent The nuclear anapole moment $\varkappa_A$ is given by (in a simple valence model)\,\cite{AnapoleFlam1984}:
\begin{equation}
\varkappa_A=1.15\times10^{-3}A^{2/3}\mu_m g_m,
\label{eq:anapole}\end{equation}
where $A$ is the number of nucleons and $\mu_m$ is the magnetic moment of the unpaired nucleon ($\mu_{p}=+2.8$, and $\mu_{n}=-1.9$). The dimensionless constant $g_m$ gives the strength of the weak interactions between the nucleons. Theoretical estimates suggest that for neutrons $g_{n}\approx-1$ and for protons $g_{p}\approx+4.5$\,\cite{AnapoleFlam1997}. From Eq.\,\eqref{eq:anapole} we see that the nuclear anapole moment scales with the number of nucleons ($\varkappa_A\propto A^{2/3}$). For this reason, the anapole moment gives the largest contribution to NSD parity-violating effects in heavy atoms\,\cite{Ginges}. Using Eq.\,\eqref{eq:anapole}, we see that the value of the anapole moment is $\varkappa_A\approx0.1-1$\,\cite{Khriplovich1980,AnapoleFlam1984}.\\
\indent The PNC matrix element is expressed in terms of a nuclear-spin-independent (NSI) and an nuclear-spin-dependent (NSD) component as follows\,\cite{Ginges}:

\begin{widetext}
	\begin{equation}\label{eq:E1PNCSISD}
		\FRME{E1_{\rm PNC}}=\FRME{E1_{\rm PNC}^{\rm (SI)}}+\FRME{E1_{\rm PNC}^{\rm (SD)}}=K_{FF'} E1_{\rm PNC} (1+r_{FF'}\varkappa),
	\end{equation}
\end{widetext}

\noindent where $K_{FF'}$ is the angular factor (using $k=1$ in Eq.\,\eqref{eq:FvsJRME}), $r_{FF'}$ is the ratio of spin-dependent to spin-independent PNC amplitudes, and $\varkappa$ is given by Eq.\,\eqref{eq:varkappa}. From Eq.\,\eqref{eq:E1PNCSISD} we see that measuring the PNC amplitudes for two different hyperfine components of a specific transition, then the value of $\varkappa$ can be expressed via the ratio of the measured amplitudes.\\
\indent The PNC rotation angle can be split into a NSI and a NSD part: 

\begin{equation}\label{eq:phiPNCSISD}
\phipnc = \varphi_{\rm SD} + \varphi_{\rm SI} = -\frac{4 \pi l}{\lambda} \:[n(\omega) - 1]\:(\R_{\rm SI} + \R_{\rm SD}).
\end{equation}
Calculated values of $E1_{\rm PNC}$, $\R$ and of the ratios $r_{FF'}$ (and thus of $\R_{\rm SI}$ and $\R_{\rm SD}$) for the various proposed transitions in Xe, Hg and I can be found in Refs.\,\cite{DzuFlaXeHg} and \cite{PNCI}.\\ 
\begin{center}
\begin{table*}[ht]
\caption{Reduced matrix elements for the $M1$ and $E1_{\rm PNC}$, and $\mathcal{R}\equiv  \text{Im}(E1_{\rm PNC})/M1$ for the proposed atomic transitions. Note that for one absorption length $\phipnc^{\rm{max}}\approx \mathcal{R}/2$.}
\label{t:pnc}
\begin{tabular}{c c c c c c  c cc c c c c}
\hline\hline\\[-1.4ex]
&$Z$ &Transition &  $\lambda$ &$\quad$&  $M1$ &$\quad$& Isotopes         &$\quad$& Im$(E1_{\rm PNC})$ &$\quad$& $\mathcal{R}$ &$\quad$\\
& & &				(nm)                        &&  ($\mu_B$) && with $I\neq0$ && $\times10^{-10}$\,$e\alpha_B$ && $\times10^{-8}$ &$\quad$\\[1.3ex]
\hline\\[-1.7ex]
                          I &53& $^2$P$_{3/2}\rightarrow ^2$P$_{1/2}$                 &               1315             && 1.15 && $^{127}$I && 0.335(67) && 0.80(16) &$\quad$\\[1.3ex] \hline\\[-1.3ex]
 \multirow{3}{*}{} && $^3$P$^{\circ}_{0}\rightarrow ^1$P$^{\circ}_{1}$  &                 609             && 0.229 && $\quad$ && 3.4(2),\,3.5(2) && 41(2),\,42(2) &$\quad$\\
 			     Hg&80&$^3$P$^{\circ}_{1}\rightarrow ^1$P$^{\circ}_{1}$   &   682              && 0.199 && \{$^{199}$Hg,\,$^{201}$Hg\} && 5.3(3),\,5.4(3) && 73(4),\,74(4) &$\quad$\\
                           && $^3$P$^{\circ}_{2}\rightarrow ^1$P$^{\circ}_{1}$  &                   997             && 0.272 && $\quad$ && 3.7(2),\,3.8(2) && 37(2),\,38(2) &$\quad$\\[1.3ex] 
\hline\\[-1.7ex]
 Xe &54& $6s\rm{}^2[3/2]^{^{\text{o}}}_{2} \rightarrow 6s^{\prime} \rm{}^2[1/2]^{^{\text{o}}}_{1}$ &        988 && 1.22 && \{$^{129}$Xe,\,$^{131}$Xe\} && 3.17(31),\,3.23(32) && 7.1(7),\,7.3(7) &$\quad$\\[1.3ex] \hline\hline
\end{tabular}
\end{table*}
\end{center}
\begin{figure}\label{fig:EnergyLevels}
	\includegraphics[width=\linewidth]{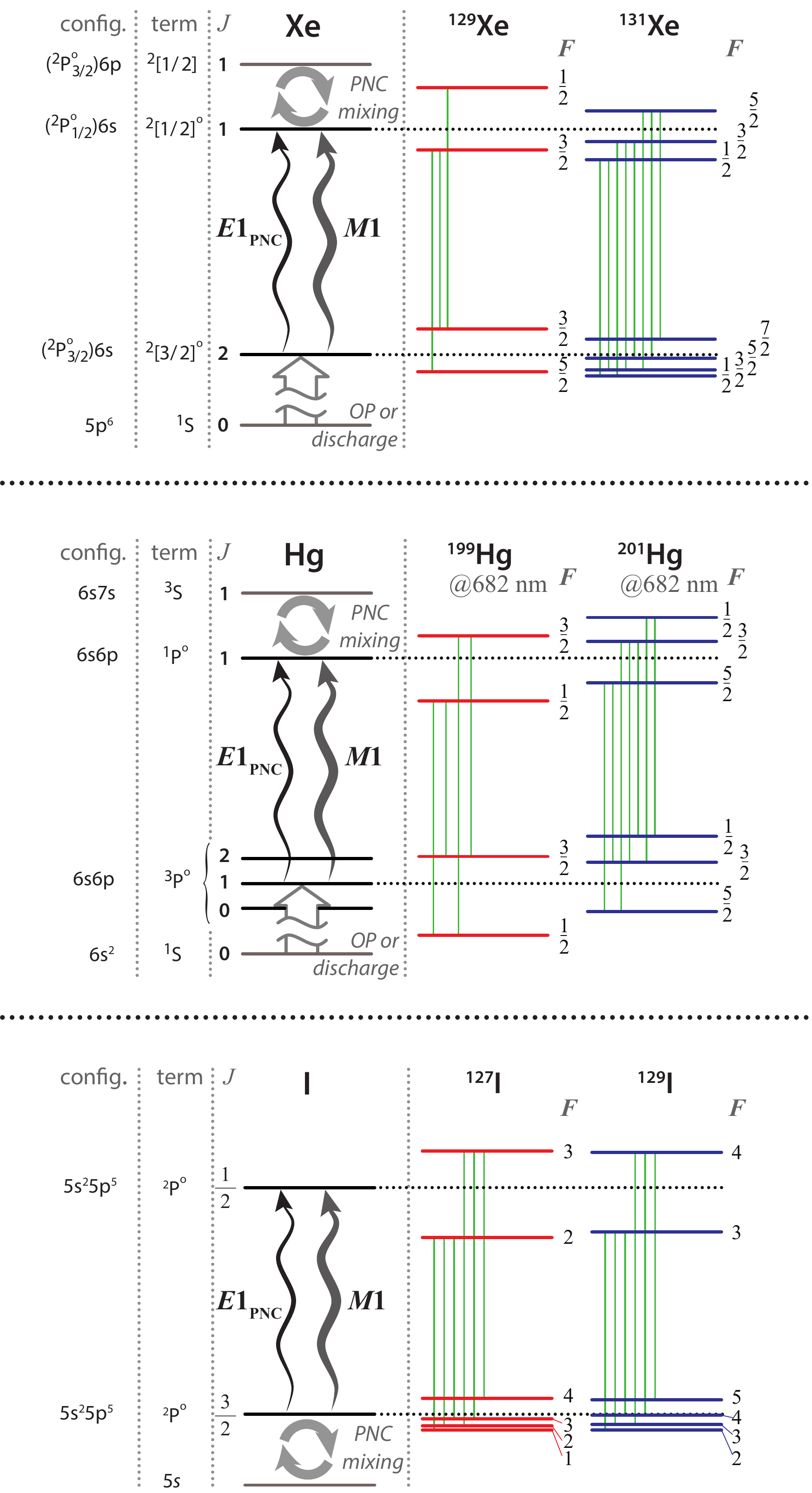}
	\caption{\label{fig:EnergyLevels} Partial energy level diagram of Xe, Hg and I (not to scale) showing the proposed $E1_{\rm PNC}$ and $M$1 transitions. In addition, the hyperfine structure levels for the odd-isotopes of Xe, Hg and I are presented. For each atomic system, indicated in green color are the individual $F\rightarrow F^{\prime}$ transitions constituting the separated hyperfine groups of Figs.\,\ref{fig:Hg},\,\ref{fig:Xe},\,\ref{fig:Iodine}.}
\end{figure}
\section{Atomic Systems \& Experimental Feasibility}\label{sec:AtomicSystems}
\subsection{PNC candidates: Xe, Hg \& I }
\indent We have identified the following favorable PNC transitions in the atomic systems of Xe, Hg and I:
(a) In metastable Xe, the $M1$ transition $(^2P^{^o}_{3/2})6s \text{ } ^2[3/2]^{^o}_{2}\rightarrow (^2P^{^o}_{1/2})6s \text{ } ^2[1/2]^{^o}_{1}$ with transition wavelength $\lambda=988$\,nm, (b) in metastable Hg, the transitions $6s6p \text{ } ^3P^{^o}_J \rightarrow 6s6p \text{ }^1P^{^o}_{1}$ at 609\,nm ($J=0$), 682\,nm ($J=1$), and 997\,nm ($J=2$), and (c) the spin-orbit transition of $^{127}$I, $^{2}$P$_{3/2}\rightarrow ^{2}$P$_{1/2}$ with transition wavelength 1315\,nm. Partial energy diagrams of the three proposed atomic systems are presented in Fig.\,\ref{fig:EnergyLevels}. \\
\indent In Bougas {\it et. al.}\,\cite{Bougas}, preliminary atomic calculations for the magnetic-dipole ($M1$) and the PNC electric-dipole ($E1_{\rm PNC}$) transition amplitudes for the proposed transitions in metastable Xe and Hg were presented (note that the simulations presented in Ref.\,\cite{Bougas} were based on these preliminary calculations). More recently, Dzuba and Flambaum\,\cite{DzuFlaXeHg}, using the configuration interaction technique, presented new calculations for the relevant transition dipole amplitudes of the proposed transitions in Xe and Hg. In particular, for the case of Hg, the spin-forbidden $M1$ transition amplitudes were overestimated in Ref.\,\cite{Bougas} and the new calculated numbers for the $M1$ dipole amplitudes, were found to be strongly suppressed. In the case of Xe the $M1$ dipole amplitude is found to be 6\% different from the initial calculation presented in Ref.\,\cite{Bougas}. In this article we use the transition amplitudes presented in Ref.\,\cite{DzuFlaXeHg} for the simulations of the expected PNC optical rotation signal under specific experimental conditions (see Section \ref{sec:IV}). In Table \ref{t:pnc} we summarize the results presented in Ref.\,\cite{DzuFlaXeHg}, along with the preliminary atomic calculations for the dipole transition amplitudes of the proposed PNC optical-rotation scheme in ground state I, as presented in Ref.\,\cite{PNCI}.\\ 
\subsection{Experimental feasibility}
\indent In the optical rotation experiments using Tl, Bi and Pb vapors, PNC optical rotation angles of $\sim$1\,$\mu$rad were measured (in the case of Tl with an experimental precision of 1\%)\,\cite{Vetter,McPherson,Meekhof1}. In order to achieve PNC rotation angles of the order of $\sim1\mu$rad, column densities of $\sim10^{18}-10^{19}$\,cm$^{-2}$ thermal atoms were required, which correspond to optical depths of 10-60. Using Eq.\,\eqref{eq:phiPNC} an estimate for the maximum expected PNC optical rotation signal can be given. Assuming a Lorentzian dispersion curve, Eq.\,\eqref{eq:phiPNC} yields $\varphi_{_\text{PNC}}\approx \mathcal{R}/2$ for one absorption length.\\
\indent The production of Xe metastable states $6s \text{ } ^2[3/2]^{^o}_{2}$ and Hg $^3P_J$ has been realized using low-pressure electrical discharge lamps \cite{Lawler,Busshian} or optical pumping \cite{Happer}, yielding steady-state densities of about $10^{12}\,\text{cm}^{-3}$, allowing column densities of about $10^{14}\,\text{cm}^{-2}$ (over a single-pass path-length of 100\,cm). Similarly, high iodine atom densities of $\sim10^{16}$\,cm$^{-3}$ have been achieved in glow discharges (requiring high precursor and carrier gas pressures). Also, the photodissociation of $\rm{I}_2$ molecules is expected to yield atomic densities of $10^{14}-10^{16}$\,cm$^{-3}$ of ground-state $^2\rm{P}_{3/2}$ iodine atoms, obtaining thus, single-pass column densities of $10^{18}\,\text{cm}^{-2}$ for an interaction path length of 100\,cm\,\cite{PNCI}.\\
\indent Given the calculated values $\R$ for Xe, Hg and I (Table\,\ref{t:pnc}), and the experimentally feasible column densities for each of the proposed atomic systems stated above, we see that single-pass PNC optical rotation angles of about $\sim10^{-11}-10^{-9}$\,rad are expected. The polarization rotation noise per unit bandwidth in a balanced polarimeter is $\sim2$\,nrad/$\sqrt{\rm Hz}$ (assuming shot-noise limited detection for a probe beam with an intensity of $\sim$10\,mW). This reasoning dictates that an additional enhancement factor ($\sim10^2-10^4$) is necessary to achieve measurable signals.\\
\indent In the following section we revisit the experimental technique proposed in Ref.\,\cite{Bougas}, and describe in detail the principles of the cavity-enhanced scheme as well as the measurement procedure.\\

\section{Cavity-Enhanced Polarimetry}\label{sec:III}
\indent In comparison to a single-pass instrument, a cavity-enhanced polarimeter introduces a phase-shift enhancement factor of $2\mathcal{F}/\pi$, where $\mathcal{F}\equiv \pi \sqrt[4]{R_t}/(1-\sqrt{R_t})$ is the finesse of the polarimeter ($R_t\!=\!R_1R_2R_3R_4$, where $R_i$ is the reflectivity of the $i$th mirror). Using high-finesse cavities, measurements with shot-noise-limited phase-shift resolution at the level of $3\times10^{-13}$\,rad have been demonstrated\,\cite{Durand}. \\
\indent In Ref.\,\cite{Bougas}, a cavity-enhanced polarimetric technique implementing signal reversals was proposed, for the enhancement and precise measurement of the PNC optical rotation angle $\phipnc$\,\eqref{eq:phiPNC}. The experimental scheme consists of a four-mirror cavity in a bow-tie configuration. A four-mirror cavity design has three main advantages over linear cavities: (a) it provides the ability of measuring simultaneously polarization effects of different symmetry under time-reversal (as in the case of magneto-optical effects and natural optical activity) without altering the apparatus during measurements, (b) it supports counter-propagating beams which give an immediate signal reversal, and (c) it avoids mechanical adjustments of possible intracavity optical elements, as in the case of two-mirror cavities used for the measurement of natural optical activity in the gas phase where intracavity quarter-wave plates needed to be modulated mechanically\,\cite{Poirson}. In this section, we present the eigenpolarization theory for the cavity-enhanced polarimeter, based on the Jones matrix calculus\,\cite{Jones, Byer, Vaccaro} and discuss in detail the principles of the proposed experimental technique.\\
\indent In the Jones matrix formalism, the effect of any optical element on the polarization state vector of the laser light is described as a linear operator, expressed by a $2\times 2$ matrix whose matrix elements are in general complex. The direct incorporation of amplitude and phase information allows for the investigation of coherent phenomena. Furthermore, since the incident CW and CCW beams will be mode-matched into the TEM$_{00}$ mode of the four-mirror cavity, we focus our analysis on the polarization properties of the longitudinal modes for either propagation direction. In addition, changes in the spatial profile of the laser beams, introduced by the intracavity elements, are neglected. The Jones matrices corresponding to each of the optical elements used in the proposed apparatus are denoted hereafter by boldface letters $\mathbf{J}$.
\begin{figure}[h!]
\begin{center}
$\begin{array}{l@{\hspace{0.21in}}r}
\multicolumn{1}{l}{\mbox{(a)}} &\\
\includegraphics[angle=0, width=0.9\linewidth]{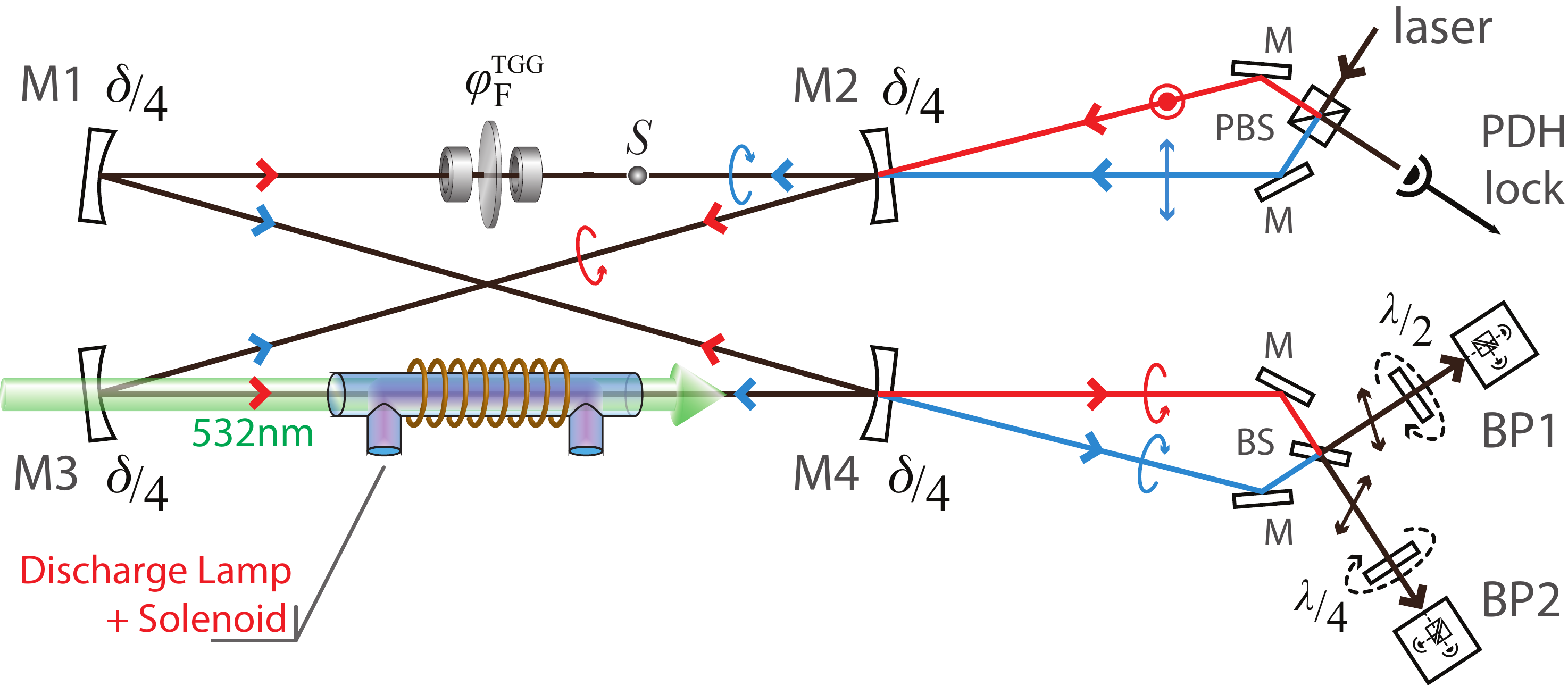}&\\[-0.1cm]
	\multicolumn{1}{l}{\mbox{(b)}} &\\ 
	\includegraphics[angle=0, width=0.9\linewidth]{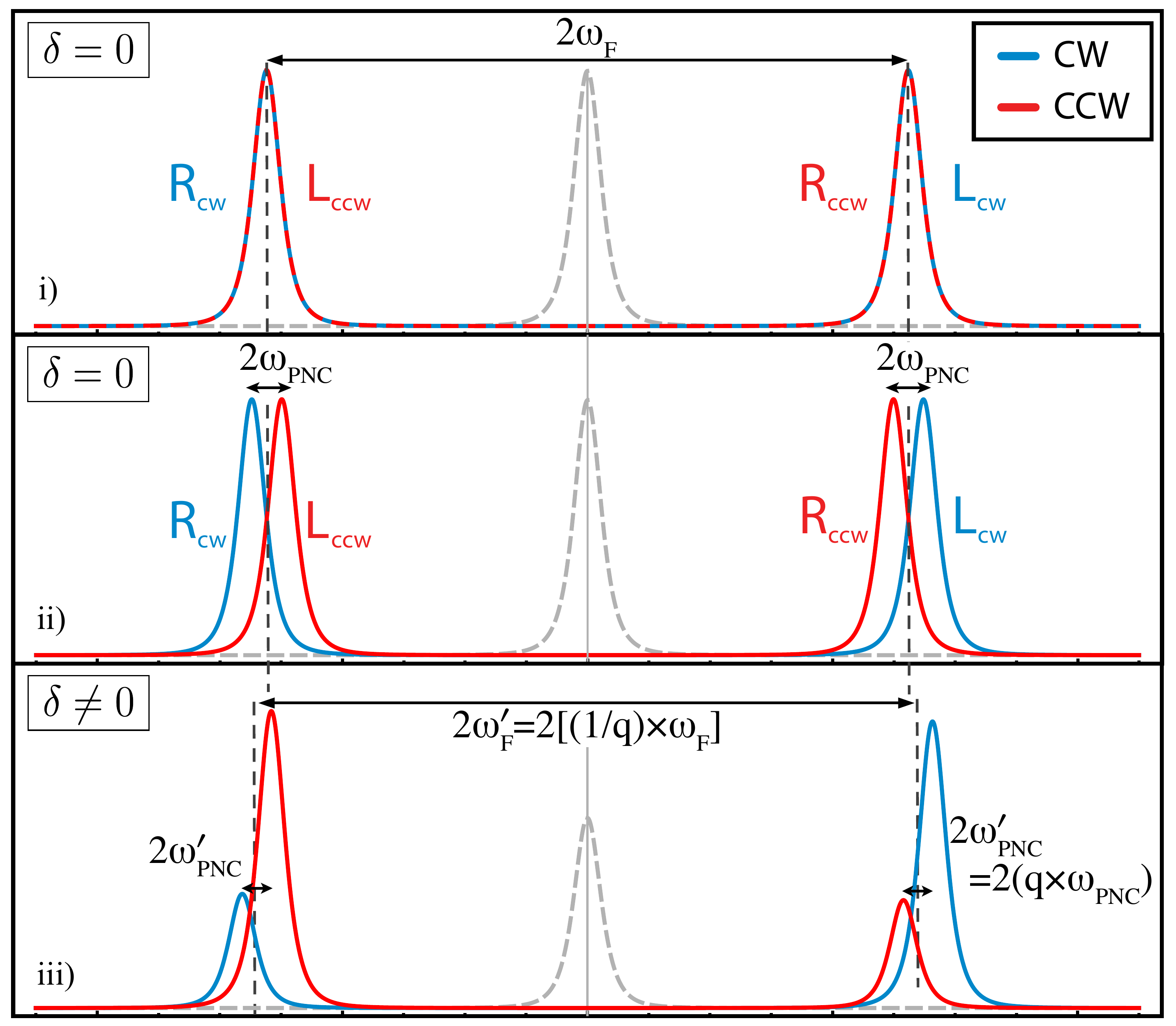}&\\[-0.1cm]
\end{array}$
\end{center}
\caption{\small{(color online). (a) Proposed experimental setup. The input laser beam is split into two parts of equal intensity and orthogonal polarizations. The laser frequency, is brought into resonance with the nearly degenerate R$_{_{\text{CW}}}$-L$_{_{\text{CCW}}}$ modes of the cavity. Upon exit from the cavity, the counterpropagating outputs are recombined into linearly polarized light, and analyzed with linear and circular balanced polarimeters (BP1 and BP2, respectively). The 532\,nm laser beam that will be used for the production of high atomic iodine densities through the photodissociation of I$_2$, is also depicted. (b) Cavity frequency polarization-spectrum: i) Faraday effect splits the cavity spectrum into $R$ and $L$ modes by $2\omega_{_\text{F}}=2\theta_{_{\text{F}}}(c/L)$ (two-fold degeneracy); ii) the PNC optical rotation splits further the CW and CCW modes by $2\omega_{_\text{PNC}}=2\phipnc(c/L)$, while the cavity modes remain circular polarization states; iii) in the presence of linear birefringence ($\delta\neq0$) the frequency splitting of the eigenmodes increases as $\omega^{\prime}_{_{\text{F}}}=1/q\,\omega_{_\text{F}}$ and the measured PNC-induced splitting is reduced $\omega^{\prime}_{_{\text{F}}}=q\,\omega_{_\text{F}}$ ($0\leqslant q\leqslant1$, see Fig.\,\ref{fig:QFactor}); the eigenmodes transform into elliptical states as observed from the different amplitudes of the output light (see text for discussion). For the simulations we assumed that the CW input beam was $p$-polarized, while the CCW beam was $s$-polarized. In i)-iii), the gray, dashed, line corresponds to the four-fold degenerate axial mode of an isotropic cavity.}}
\label{fig:PNCexp}
\end{figure}
\begin{figure}[h!]
\includegraphics[angle=0, width=1.\linewidth]{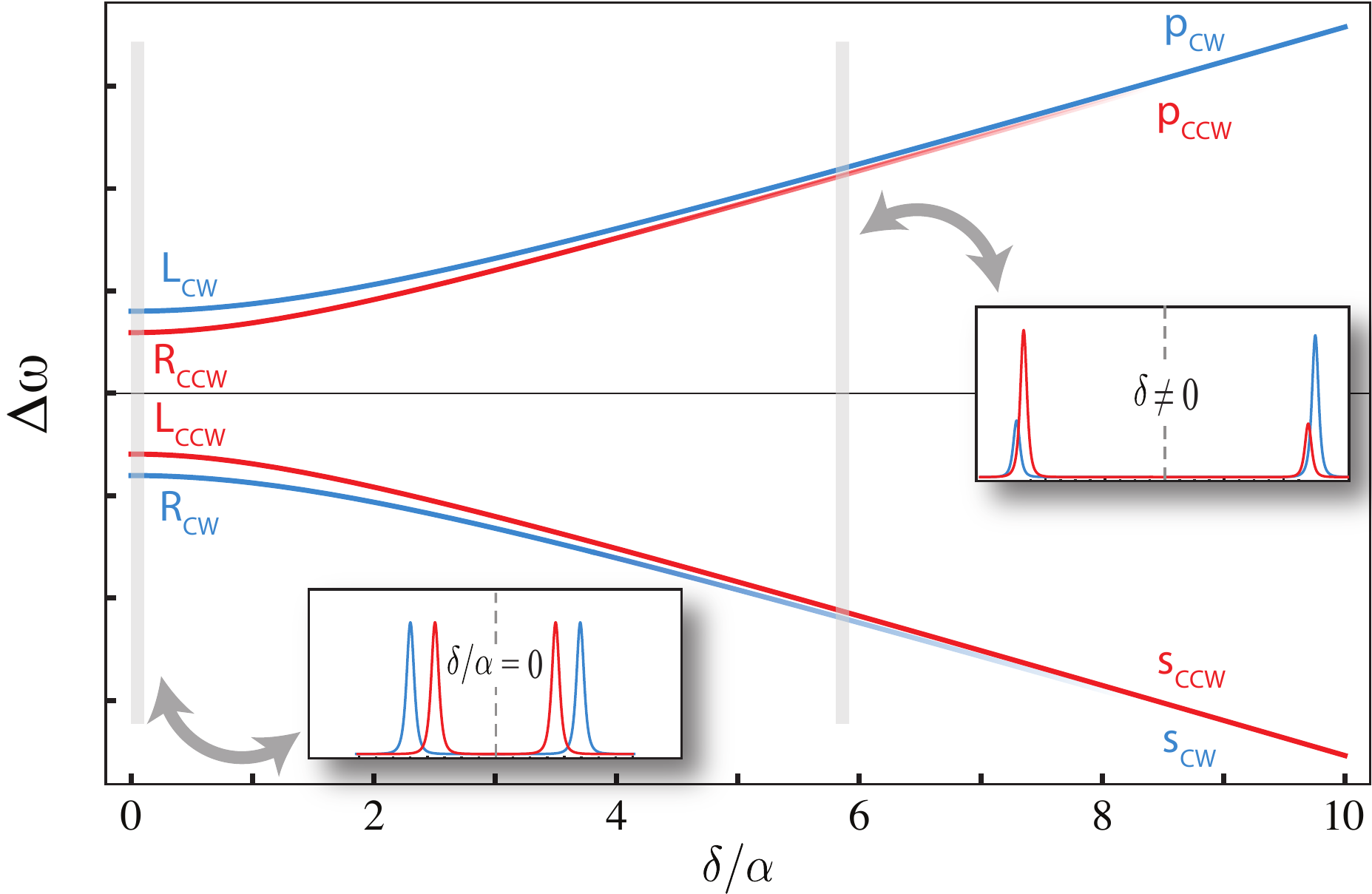}
\caption{\small{(color online) The presence of linear birefringence $\delta$ prevents the enhancement of the PNC optical rotation $\phipnc$. The resonance peaks of the cavity eigenpolarization frequency spectrum are presented as a function of the ratio $\delta/\alpha$. As $\delta$ increases, and therefore the magnitude of the total cavity anisotropies increases, the frequency difference between the respective CW (or CCW) R-L modes increases by $1/q$, while the PNC-induced frequency splitting (exaggerated here for clarity) is decreasing by q (Eq.\,\eqref{eq:QFactor}).}}
\label{fig:QFactor}
\end{figure}
\subsection{Jones Matrices for Polarization Optics}
The Jones matrix for reflection is the same for CW and CCW propagation and is given by:
\begin{equation}
\mathbf{J}_{{\rm M}_i}(\delta_i)=\sqrt{R_i} \left(\begin{array}{cc}-e^{i\delta_i/2} & 0 \\0 & e^{-i\delta_i/2} \end{array}\right),
\end{equation}
where the index $i$ ranges from 1 to 4. We assume that the Fresnel amplitude reflection coefficients for the \textit{s} and \textit{p} polarizations are equal in magnitude (an assumption expressed by the common factor $\sqrt{R_i}$), which is a good approximation for near-normal angle-of-incidence reflections, as in the case of a bow-tie cavity. The differential $s$-$p$ phase shift $\delta_{i}=\delta_p-\delta_s$, represents the linear birefringence obtained upon mirror reflection. For non-normal incidence, these $s$-$p$ phase shifts can be of the order of $10^{-3}$ rad, while for normal incidence of the order of $10^{-5}$ to $10^{-6}$ rad at a specific design wavelength (for gyroquality super-mirrors  at normal incidence, the linear birefringences can be as low as $\sim\!0.1 \mu$rad)\,\cite{Hall}.\\
\indent In the presence of a longitudinal magnetic field a medium becomes circular birefringent, an effect otherwise know as the Faraday effect\,\cite{BudkerRMP}. The Faraday optical rotation is expressed as: $\theta_{_{\text{F}}}=V\text{B}l$, where B is the magnetic field strength along the direction of light propagation, $l$ is the pathlength of interaction, and $V$ is the Verdet constant of the medium. The Jones matrix for the Faraday rotation is an SU(2) rotation matrix with argument $\theta_{_{\text{F}}}$:
\begin{equation}
\mathbf{J}_{_{\text{F}}}(\theta_{_{\text{F}}})=\left(\begin{array}{cc} \cos\theta_{_{\text{F}}}& -\sin\theta_{_{\text{F}}} \\ \sin\theta_{_{\text{F}}} & \cos\theta_{_{\text{F}}} \end{array}\right).
\end{equation}
Note that the physical direction of the polarization rotation is defined by the magnetic field orientation. Due to the non-reciprocal nature of the Faraday effect, when either the magnetic field or the direction of propagation of the light reverses, the sign of rotation reverses (in the light-frame). Thus, for the CCW propagation, the Faraday rotation will be $\theta^{\rm{ccw}}_{_{\rm{F}}} \to -\theta^{\rm{cw}}_{_{\text{F}}}$. As we shall see, this directional symmetry breaking, induced by the Faraday effect, is essential to our signal reversals.\\
\indent The Jones matrix representing the PNC optical rotation, will also be that of an SU(2) rotation matrix with argument $\phipnc$: 
\begin{equation}
\mathbf{J}_{_{\text{PNC}}}(\phipnc)=\left(\begin{array}{cc} \cos\phipnc& -\sin\phipnc \\ \sin\phipnc & \cos\phipnc \end{array}\right),
\end{equation}
where $\phipnc$ is given by Eq.\,\eqref{eq:phiPNC}. The PNC optical rotation, being a pseudoscalar quantity, is odd under parity transformations \emph{and} even under time-reversal transformations. Therefore, the Jones matrix describing PNC optical rotation will be the same for both CW and CCW propagation directions, $\phipnc^{\rm{cw}}=\phipnc^{\rm{ccw}}$. \\
\indent Finally, anisotropies such as imperfections of transmission optics, thermal or stress induced birefringences, and stray magnetic fields, can be described as linear birefringent optical elements. The Jones matrix for a general linear wave-retarder, which introduces a differential phase shift $\delta^{\prime}$, and whose ``fast axis'' is oriented at an angle $\theta$ with respect to the x-axis, is given by:
\begin{equation}
\mathbf{J}(\theta,\delta^{\prime})=S(\theta)\times \left(\begin{array}{cc} e^{i\delta^{\prime}/2} & 0 \\0 & e^{-i\delta^{\prime}/2} \end{array}\right) \times S(-\theta),
\end{equation}
where $S(\theta)$ describes a general SU(2) rotation matrix. Reversing the direction of propagation (in the light frame), reverses the sign of the angle $\theta$ which specifies the orientation of the retardation axes. For the mirror-reflection linear birefringence, we used $\mathbf{J}(\theta,\delta^{\prime})$ for $\theta=0$. Note that the eigenvectors of the $\mathbf{J}(\theta,\delta^{\prime})$ are linear polarization states.\\
\subsection{CW and CCW round trip matrices} 
\indent The round-trip Jones matrices for the CW (CCW) propagation are obtained by the ordered multiplication of the Jones matrices representing the optical elements. A convenient starting point for the analysis is the point labeled $S$ in Fig. \ref{fig:PNCexp}, from which the different propagation directions are defined. The round trip Jones matrices are given by:
\begin{equation}
\mathbf{J}^{_{\text{CW}}}\!=\!\mathbf{J}_{\rm M_2}(^\delta\!/\!_4)\!\cdot\! \mathbf{J}_{\rm M_3}(^\delta\!/\!_4)\!\cdot\! \mathbf{J}(\varphi_{_\text{PNC}})\!\cdot\! \mathbf{J}(\theta_{\text{F}}) \!\cdot\! \mathbf{J}_{\rm M_4}(^\delta\!/\!_4) \!\cdot\! \mathbf{J}_{\rm M_1}(^\delta\!/\!_4)
\label{eq:RTcw}
\end{equation}
for the CW propagation path, and 
\begin{equation}
\mathbf{J}^{_{\text{CCW}}}\!=\!\mathbf{J}_{\rm M_2}(^\delta\!/\!_4)\!\cdot\! \mathbf{J}_{\rm M_3}(^\delta\!/\!_4)\!\cdot\! \mathbf{J}(-\theta_{_\text{F}})\!\cdot\! \mathbf{J}(\varphi_{_\text{PNC}}) \!\cdot\! \mathbf{J}_{\rm M_4}(^\delta\!/\!_4) \!\cdot\! \mathbf{J}_{\rm M_1}(^\delta\!/\!_4) 
\label{eq:RTccw}
\end{equation}
for the CCW propagation path. Here, we define $\delta$ as the total single-pass linear birefringence. Note that by reversing the order of the individual operators and changing the sign of each Faraday rotation angle for the CW (CCW) path produces the CCW (CW) path (if an additional linear birefringent element is present, then the sign of its respective orientation angle should be also reversed so as to obtain the CCW propagation matrix).\\
\indent The Jones matrices for the Faraday rotation and the PNC rotation are commutable, a property that reflects the fact that rotations about the same axis are additive ($\mathbf{J}(\varphi_{_\text{PNC}})\!\cdot\! \mathbf{J}(\theta_{\text{F}})=\mathbf{J}(\varphi_{_\text{PNC}}+\theta_{\text{F}})$). Therefore, the total single-pass optical rotation is different for the CW and CCW counterpropagating beams: 
\begin{equation}\alpha_{_{\text{CW}}}\!=\!\theta_{_{\text{F}}}+\phipnc\quad \text{and} \quad  \alpha_{_{\text{CCW}}}\!=-\!\theta_{_{\text{F}}}+\phipnc.
\label{eq:AlphaCWCCW}
\end{equation} 
This directional symmetry breaking is key for distinguishing the PNC and Faraday type optical rotation and thus for the sensitive measurement of the PNC optical rotation angle.\\
\indent Re-writing Eq.\,\eqref{eq:RTcw} and Eq.\,\eqref{eq:RTccw} in a compact form, we have: 
\begin{align}
\mathbf{J}^{_{\text{CW}}}& =  R^2\!\cdot\! \mathbf{J}(0,\delta/2)\!\cdot\! \mathbf{J}(\alpha_{_\text{CW}})\!\cdot\! \mathbf{J}(0,\delta/2), \label{eq:Jcw} \\
\mathbf{J}^{_{\text{CCW}}}& =  R^2\!\cdot\! \mathbf{J}(0,\delta/2) \!\cdot\! \mathbf{J}(\alpha_{_\text{CCW}})\!\cdot\! \mathbf{J}(0,\delta/2), \label{eq:Jccw}
\end{align}
where we omit the mirror index under the assumption that all four mirrors have the same reflectivity and linear birefringence.\\
\subsection{Frequencies and polarizations of cavity spectrum}
\indent The allowed polarizations of the cavity modes, along with their respective frequencies, are determined by the anisotropies of the cavity. Using the explicit form of the transfer matrices for CW and CCW propagation (Eq. \eqref{eq:Jcw},\,\eqref{eq:Jccw}) we can obtain the eigensystem for both directions as a function of the anisotropy parameters ($\theta_{_\text{F}}$, $\varphi_{_\text{PNC}}$, and $\delta$)\,\cite{Byer}. For the following discussion, we set $R=1$, as we are interested only in the properties of the frequency spectrum of the optical resonator.\\
\indent The matrices $\mathbf{J}^{_{\text{CW}}}$ and $\mathbf{J}^{_{\text{CCW}}}$ are unitary matrices, of rank two. Therefore, each matrix has two eigenvalues and two eigenvectors; the eigenvectors $\nu_{\pm}$ are generally complex, orthogonal vectors, and represent the eigenpolarizations of each cavity mode. The eigenvalues can be written in the form $\lambda_{\pm}=e^{\pm i\Phi}$. The phase of each eigenvalue is the round-trip optical phase shift obtained during light propagation, and therefore yields the frequency splittings of the eigenmodes. \\
\indent  In the simple case of an isotropic cavity ($\alpha=0$ and $\delta=0$), the four eigenmodes are degenerate and any polarization state can couple into the cavity ($\mathbf{J}^{_{\text{CW}}}$ and $\mathbf{J}^{_{\text{CCW}}}$ become proportional to the identity matrix for $\alpha=0$ and $\delta=0$). The introduction of anisotropies lifts this four-fold degeneracy. Therefore, in the most general case, the spectrum of the cavity is represented by four non-degenerate modes of elliptical polarization, whose frequencies lie above and below the degenerate frequency of the isotropic case. We examine three cases.\\
\indent i) \emph{$\theta_{_\text{F}}\neq0$, $\phipnc=0$, and $\delta=0$ : } The Jones matrices for CW and CCW become:
\begin{equation}
\mathbf{J}^{_{\text{CW}}}=\mathbf{J}_{_{\text{F}}}(\theta_{_{\text{F}}}) \quad\text{and}\quad  \mathbf{J}^{_{\text{CCW}}}=\mathbf{J}_{_{\text{F}}}(-\theta_{_{\text{F}}}).
\end{equation}
\indent It is easy to verify that the allowed eigenpolarizations of a rotation matrix are circular polarization states. Therefore, in the presence of single pass Faraday rotation $\theta_{_\text{F}}$, the spectrum splits into right circular (RCP) and left circular (LCP) polarization modes; the frequency splitting is equal to $2\omega_{_\text{F}}=2\theta_{_\text{F}}(c/L)$, where c is the speed of light and L is the round-trip cavity length. The non-reciprocal nature of the Faraday effect, embedded in the change of sign of the Faraday rotation when the direction of propagation is reversed, is directly reflected in the frequency spectrum of the cavity. The $R_{_{\text{CW}}}$ mode is degenerate with the $L_{_{\text{CCW}}}$ while the $R_{_{\text{CCW}}}$ mode is degenerate with the $L_{_{\text{CW}}}$ mode (see Fig. \ref{fig:PNCexp}(b), (i)).\\ 
\indent ii) \emph{$\theta_{_\text{F}}, \phipnc\neq0$, and $\delta=0$ : }
For single-pass rotations $\varphi_{_\text{PNC}}$ and $\theta_{_\text{F}}$, and in the absence of any linear birefringence ($\delta=0$), the round-trip matrices for CW and CCW correspond to rotation matrices with arguments $\alpha_{_{\text{CW}}}$ and $\alpha_{_{\text{CCW}}}$ (Eq. \eqref{eq:AlphaCWCCW}):
\begin{align}
\mathbf{J}^{_{\text{CW}}}=\mathbf{J}(\alpha_{_{\text{CW}}}) \quad\text{and}\quad  \mathbf{J}^{_{\text{CCW}}}=\mathbf{J}(\alpha_{_{\text{CCW}}}).
\end{align}
The eigenpolarizations remain circular polarization states for both propagation directions, since the transfer matrices are simply rotation matrices. Their respective eigenvalues are $\lambda^{\pm}_{_\text{CW}}=e^{\pm i\alpha_{_{\rm{cw}}}}$ and $\lambda^{\pm}_{_\text{CCW}}=e^{\pm i\alpha_{_{\text{ccw}}}}$. The difference in rotation (Eq. \eqref{eq:AlphaCWCCW}), results in splitting the CW and CCW modes by $2\omega_{_\text{PNC}}=2\varphi_{_\text{PNC}}(c/L)$, yielding the four-mode structure depicted in Fig. \ref{fig:PNCexp}(b), case (ii).\\
\indent iii) \emph{$\theta_{_{\text{F}}}$,\, $\phipnc$, and $\delta \neq 0$ : }
Linear birefringence prevents the enhancement of circular birefringence through the transformation of a linearly polarized beam into a circular one. If however, a large circular birefringence is induced, then the effects of linear birefringence wil be averaged out\,\cite{Bougas},\cite{Vaccaro}. Using the general form of the CW and CCW matrices (Eq.\,\eqref{eq:Jcw} and \eqref{eq:Jccw}) we demonstrate how the extraction of $\phipnc$ is affected in the presence of $\delta$. Expanding Eq. \eqref{eq:Jcw} and \eqref{eq:Jccw}, we get:
\begin{align}
\mathbf{J}^{_{\text{CW}}}&=\left(\begin{array}{cc} e^{\frac{i\delta}{2}}\cos(\theta_{_{\text{F}}}\!+\phipnc)& -\sin(\theta_{_{\text{F}}}\!+\phipnc) \\ \sin(\theta_{_{\text{F}}}\!+\phipnc) &e^{-\frac{i\delta}{2}} \cos(\theta_{_{\text{F}}}\!+\phipnc) \end{array}\right)\\
\mathbf{J}^{_{\text{CCW}}}&=\left(\begin{array}{cc} e^{\frac{i\delta}{2}}\cos(\theta_{_{\text{F}}}\!-\phipnc)& \sin(\theta_{_{\text{F}}}\!-\phipnc) \\ -\sin(\theta_{_{\text{F}}}\!-\phipnc) &e^{-\frac{i\delta}{2}} \cos(\theta_{_{\text{F}}}\!-\phipnc) \end{array}\right).
\end{align}
The eigenvalues and eigenvectors are:
\begin{align}
\lambda^{\pm}_{_\text{cw}}&=\cos\alpha_{_{\text{cw}}}\cos\frac{\delta}{2} \mp i\sqrt{1-\cos^2\alpha_{_{\text{cw}}} \cos^2\frac{\delta}{2} }\nonumber\\
\quad &  \nonumber \\
\nu^{\pm}_{_\text{cw}}\!&\propto\! \left(\! \begin{array}{ccc}
\csc\alpha_{_{\text{cw}}}\left(\!\cos\alpha_{_{\text{cw}}} \sin\frac{\delta}{2}\! \mp \! \sqrt{1-\cos^2\alpha_{_{\text{cw}}} \cos^2\frac{\delta}{2}\!}\right)  \\
\quad \\
-i \\
\end{array}\! \right)\!.
\label{eq:eigensystemCW}
\end{align}
for the CW transfer matrix, and
\begin{align}
\lambda^{\pm}_{_\text{ccw}}&=\cos\alpha_{_{\text{ccw}}}\cos\frac{\delta}{2} \mp i\sqrt{1-\cos^2\alpha_{_{\text{ccw}}} \cos^2\frac{\delta}{2} }\nonumber \\
\quad &  \nonumber \\
\nu^{\pm}_{_\text{ccw}}\!&\propto\!\left(\! \begin{array}{ccc}\!
\csc\alpha_{_{\text{ccw}}}\!\left(\!\cos\alpha_{_{\text{ccw}}} \sin\frac{\delta}{2}\! \mp \! \sqrt{1-\cos^2\alpha_{_{\text{ccw}}} \cos^2\frac{\delta}{2}\!}\right)  \\
\quad \\
i \\
\end{array}\! \right)\!.
\label{eq:eigensystemCCW}
\end{align}
for the CCW transfer matrix. We see that in the most general case the polarizations eigenstates, for both the CW and CCW modes, are represented by orthogonal ellipses and their frequency splitting is proportional to $\Gamma=\cos^{-1}[ \cos\alpha\cos(\delta/2)]$. \\ 
\indent Linear birefringence $\delta$ prevents the effective amplification of circular birefringence $\alpha$ by transforming the cavity modes into elliptical polarization states. Therefore, the measurement of $\varphi_{\text{PNC}}$ in the presence of linear birefringence will be reduced to: $\varphi^{\prime}_{\text{PNC}}=q\,\varphi_{\text{PNC}}$, where $q$ ($0\leqslant q\leqslant1$) is the reduction factor. From Eq. \eqref{eq:eigensystemCW} and \eqref{eq:eigensystemCCW}, we obtain the form of the reduction factor for $\varphi_{_\text{PNC}}\ll1$:
\begin{equation}\label{eq:QFactor}
q=\frac{\Gamma-\Gamma|_{\phipnc=0}}{\phipnc}=\frac{\cos\delta/2\sin\theta_{_\text{F}}}{\sqrt{1-\cos^2\delta/2\cos^2\theta_{_\text{F}}}}+\mathcal{O}(\phipnc).
\end{equation}
\indent In Fig.\,\ref{fig:QFactor}, we investigate the effect of the linear birefringence $\delta$ as a function of the ratio of the total linear birefringence anisotropy over the total circular birefringence, $\delta/\alpha$. The introduction of this extra anisotropy ($\delta$) will increase the frequency splitting of the modes for each sense of propagation. This effective increase in frequency is inversely proportional to and equal in magnitude with the simultaneous decrease of the PNC-induced splitting, i.e.~$2\Gamma/(2\Gamma|_{\delta=0})\equiv1/q$ (Fig. \ref{fig:QFactor}). Fig.\,\ref{fig:PNCexp}\,(b) case (iii), and Fig.\,\ref{fig:QFactor}, show simulations based on Eq.\,\eqref{eq:eigensystemCW} and Eq.\,\eqref{eq:eigensystemCCW}, which demonstrate how the presence of linear birefringence prohibits the enhancement of circular birefringence, as the PNC-induced mode splitting vanishes for large $\delta/\alpha$. Note that the cavity's eigenpolarization-modes become more elliptical with increasing $\delta$; the input beams are linearly polarized, and therefore the induced ellipticity is depicted on the different intensity amplitudes of the cavity (output) modes. Therefore, to ensure $q\cong1$, one must satisfy $\alpha\gg\delta$ (see also the relevant discussion in ref. \cite{Bougas}).\\
\subsection{Principles of the Measurement} 
\indent The principles of the measurement have been described previously in Ref.\,\cite{Bougas} and are briefly discussed here. We will assume for the following discussion a bow-tie four-mirror cavity with round-trip cavity length $L=7.5$\,m.\\
\indent A laser beam is split into two beams of equal intensity and orthogonal linear polarizations. The $s$-polarized laser beam is locked to the R$_{_{\text{CW}}}$ mode and frequency-locked using the Pound-Drever-Hall (PDH) scheme \cite{PDH}. Note that alternative locking schemes have also demonstrated shot-noise-limited phase-shift measurements (see Ref.\,\cite{Durand} and references therein). The PNC-related mode splitting is equal to $2\omega_{_{\text{PNC}}}=2\phipnc c/L$. For the different values of $\mathcal{R}$ presented in Table\,\ref{t:pnc}, we get $\omega^{\rm max}_{_{\text{PNC}}}\sim150$\,mHz-15\,Hz. The PNC-induced mode splitting is much smaller than the cavity linewidth $\Delta\omega_{\rm cav.}$ (for $L=7.5$\,m and $\mathcal{F}\sim1.5\times10^4$, $\Delta\omega_{\rm cav.}=2\pi\times2.5$\,kHz). Therefore the $p$-polarized laser beam excites the nearly degenerate L$_{_{\text{CCW}}}$ mode (see Fig.\,\ref{fig:PNCexp}\,(b)). The spatial recombination of the R$_{_{\text{CW}}}$ and L$_{_{\text{CCW}}}$ output beams produces a linearly polarized beam rotated by $N\phipnc$, where $N$ is the average number of round-trip cavity passes. The rotation angle $N\phipnc$ will be measured with a balanced polarimeter. Note that the spatial recombination of the two output beams is expected to be a source of depolarization, for which the signal needs to be corrected. Therefore, we propose the use of two separate balanced polarimeters, implementing rotating half-wave and quarter-wave plates respectively, yielding the complete set of Stokes parameters of the output recombined light (see Ref. \cite{Berry}). \\
\indent Observe that bringing the CW and CCW beams into resonance with the R$_{_{\text{CCW}}}$-L$_{_{\text{CW}}}$ mode pair, the recombination of the exit beams will give now a signal output of $-N\phipnc$, yielding thus a net difference in polarization rotation of $2N\phipnc$. This is accomplished through the use of two signal reversals. First, the frequency of the laser can be brought into resonance with the R$_{_{\text{CCW}}}$-L$_{_{\text{CW}}}$ mode pair with the use of an acoustic-optic-modulator (AOM). Second, reversing the magnetic field is equivalent to the interchange of the CW and CCW beams, and thus the laser will couple to the R$_{_{\text{CCW}}}$-L$_{_{\text{CW}}}$ mode pair. These two novel signal reversals allow for the absolute measurement of the PNC optical rotation, avoiding the need for cell removal, as was required in previous PNC optical rotation experiments. Additionally, these reversals can be performed at high frequencies of $\sim1$\,kHz, allowing a sufficient subtraction of experimental drifts. Note, that the frequency of the reversals are constrained by the photon lifetime inside the cavity (for $R=R_1R_2R_3R_4=0.9999^4$ then $\tau_{\rm{photon}}=L/(c|{\rm ln}R|)\sim63\,\mu\rm{sec}$). \\
\indent In the previous subsections, we saw that a linear birefringence can suppress the enhancement of the PNC optical rotation. In general, linear birefringences originating from mirror-reflection phase-shifts, thermal and/or stress-induced birefringences, are expected to be $\sim\!10^{-3}$\,rad (per single-pass per reflection or transmission). Inducing a large circular birefringence protects the coherent accumulation of the PNC optical rotation inside the cavity. The circular birefringence can be induced using the Faraday effect of the proposed transitions themselves. Theoretical calculations for the Faraday effect on the $M1$ transitions under consideration, and the proposed column densities, yield $\theta_{_{\text{F}}}\!\sim\!10^{-3}$ rad for a $200$\,G magnetic field\,\cite{Sand,Vet}. An alternative is the use of an anti-reflection (AR) coated high-Verdet glass window inside the cavity, a dense flint glass for example. A Terbium Gallium Garnet (TGG) crystal has a Verdet constant of V$\sim\,$45\,$\mu$rad\,G$^{-1}$\,cm$^{-1}$ with losses of $\sim10^{-4}/$\,mm$^{-1}$ at 1064\,nm. For a 1\,mm crystal thickness and magnetic fields of 3000\,G one obtains $\theta_{_\text{F}}\!\sim\!13.5$\,mrad, ensuring that $\alpha\gtrsim10\delta$ for which the depolarization factor $q\gtrsim0.9993$ (see also discussion in Ref.\,\cite{Bougas}). Finally, note that in the case of large linear birefringences, a compensator (for example, a MgF$_2$ thin glass) with a high antireflection coating can be placed appropriately to reduce the cavity's total linear birefringence, and therefore to satisfy the condition $\alpha\gtrsim10\delta$. \\
\indent As a final remark, note that the metastable Xe and Hg are produced in a discharge lamp or via optical pumping, or in the case of I, from molecular photodissociation, and can thus be ``switched'' on and off. This gives us an additional subtraction procedure which allows for the real-time investigation of the ``empty" cavity and thus of possible experimental errors.\\
\section{Theoretical Simulations}\label{sec:IV}
\subsection{Optical Absorption Length}
\begin{figure}
\centering
\includegraphics[width=\linewidth]{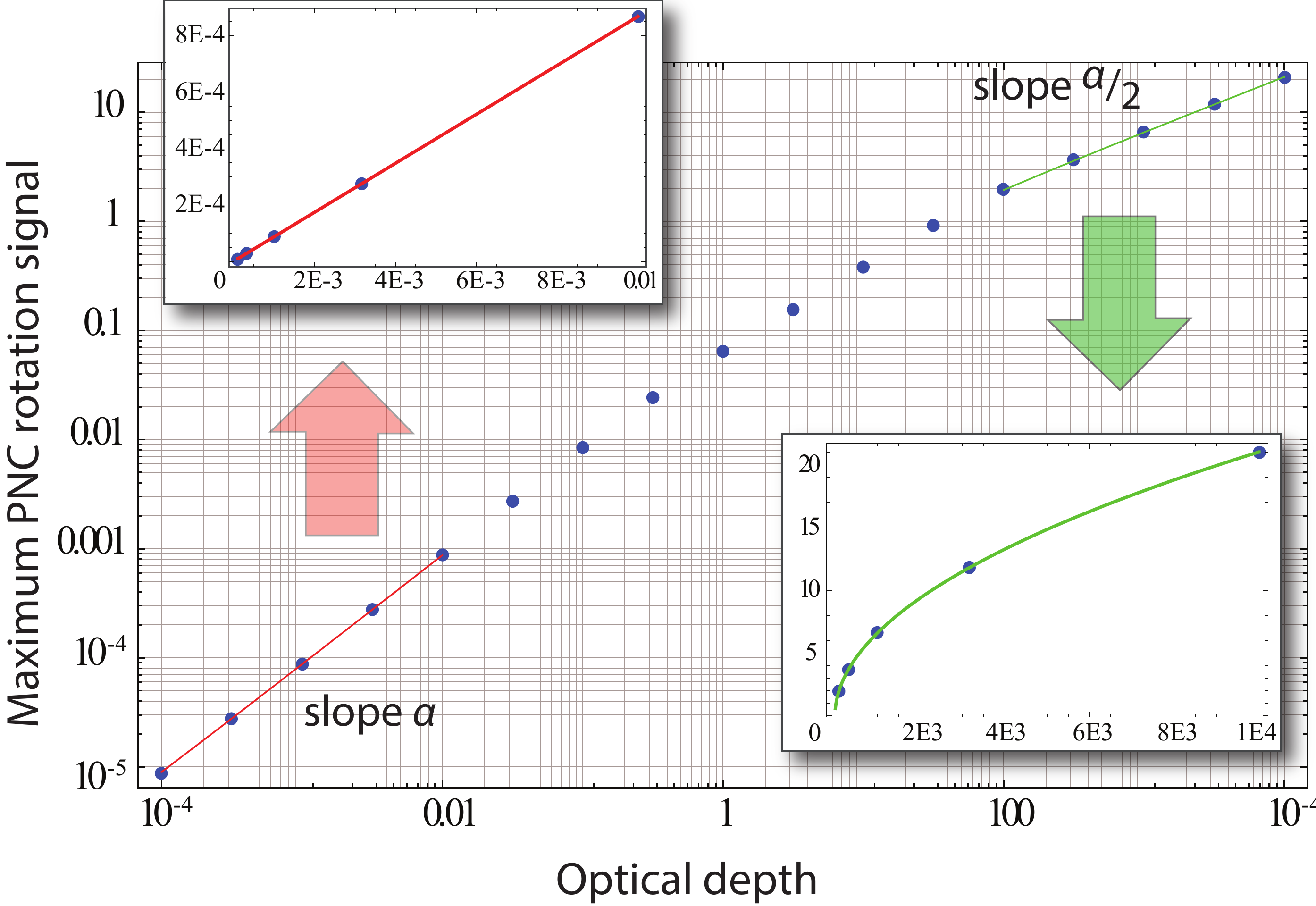}
\caption{\small{The PNC optical rotation signal is proportional to the product $\phipnc(\omega)\times T(\omega)$ (Eq.\,\eqref{eq:signal}). Assuming Voigt line shape profiles, the maximum PNC rotation signal is plotted as a function of the resonant optical depth (OD). We demonstrate that the signal scales linearly with OD when the vapor is optically thin, and continues to increase with a square root dependence as the vapor becomes thicker. The $y$-axis is given in units of $\mu$rad, and we assumed $\mathcal{R}=14\times10^{-8}$.}}
\label{fig:SvsOD}
\end{figure}

Upon exiting from the cavity, the recombined laser beams will be analyzed by a balanced polarimeter. The detected signals will be of the form:
\begin{equation}\label{eq:signal}
S=2 N\phipnc(\omega) \times T(\omega),
\end{equation}
where $N$ is the average number of round-trip cavity passes, $\phipnc(\omega)$ denotes the dispersive line-shape of the PNC optical rotation, and $T(\omega)$ is the transmission of the light beam through the vapor, which is governed by the Beer-Lambert law\,\cite{Sob}:
\begin{equation}\label{eq:Tr}
T(\w) = \frac{I(\w)}{I_{\rm o}}= e^{-A(\w)}\equiv e^{-\rho\sigma(\w) l}.
\end{equation}
Here, $A(\omega)$ is defined as the absorptivity in terms of the interaction path-length $l$, the number density of the atoms $\rho$, and the absorption cross section $\sigma(\omega)$, which is a function of the optical frequency. $I_{\rm o}$ is the intensity of the incident laser beam. \\
\indent The absorption cross section, $\sigma$, is given by the expression:\\
\begin{equation}\label{eq:sigma}
\sigma(\w) = \sigma_{\rm o}\sum_i\sum_{F,F'} b_i\:C_{FF'} \V''_{FF',i}(\w),
\end{equation}
where, $b_i$ is the abundance of isotope $i$ (see Appendix \ref{app:n}), the $C_{FF'}$ are geometry factors (Eq.\,\eqref{eq:CFFQ}) and $\V''_{FF'}$ is the absorptive part of the Voigt profile (given in Eq.\,\eqref{eq:voigtA}). In the equation above, the \emph{integrated absorption cross section}, $\sigma_{\rm o}$, is: 
\begin{equation}\label{eq:sigmao}
\sigma_{\rm o} = \frac{\pi\mo\w_{JJ'}}{\hbar~c}\: \frac{1}{2 J+1}\: \frac{M1^2}{3}.
\end{equation}
\noindent Note that, $\sum_i\sum_{F,F'} b_i\:C_{FF'} = 1$ and, since $\int_{0}^{\infty}\V''(\w) d\w = 1$, we get $\int_{0}^{\infty}\sigma(\w) d\w = \sigma_{\rm o}$, hence $\sigma_{\rm o}$ is justifying its name. In addition, $\sigma_{\rm o}$ does not have units of area.\\
\indent The extremely long effective path-lengths, that can be realized in stable high-finesse optical cavities, lead to large effective resonant absorption lengths. In Fig.\,\ref{fig:SvsOD} we present calculations for the maximum PNC optical rotation signal expected, as a function of the resonant absorption optical lengths ($l_0$). The PNC optical rotation signal is proportional to the product $\phipnc(\omega)\times T(\omega)$ (Eq.\,\eqref{eq:signal}), i.e. proportional to the product of a dispersive line-shape profile times an absorption line-shape profile. For resonant optical depths $l_0\ll1$, the maximum PNC optical rotation angle increases linearly with increasing column densities, i.e. $\phipnc^{\max} \propto \rho l$ (where $\rho$ is the density and $\rho l$ is column density of the vapor), as seen in the first inset of Fig.\,\ref{fig:SvsOD}. For optical depths $l_0\gg1$, the vapor is optically thick near the line center where $\phipnc$ is largest and can no longer be observed. The effective maximal rotation angle is shifted further off resonance as $\sqrt{\rho l}$, and $\phipnc^{\max } \propto \sqrt{\rho l}$, as can be shown by maximizing the product of dispersion and transmission. Therefore, the rotation angle can still be increased with increasing column density for $l_0\gg1$, only with a rate slower than linear (see second inset of Fig.\,\ref{fig:SvsOD}). For example, in the case of Tl and Pb, vapor densities producing values of $A(\omega)\sim10-60$ absorption lengths at the line center of the $M1$ transitions were realized, obtaining thus PNC rotation angles of about $10^{-6}\,\mu$rad at the dispersion peaks.\\

\subsection{PNC Optical Rotation Simulations}\label{subsec:Sims}
\indent In this section we present theoretical simulations of the PNC optical rotation signals for the proposed transitions in Xe, Hg and I, where we explore a range of experimentally feasible parameters. We assume a four-mirror bow-tie cavity of round-trip cavity length $L=7.5$\,m (free spectral range FSR\,$=\,40$\,MHz), each mirror having a reflectivity of $R=99.99$\% (enhancement factor $N$\,$\sim$\,$10^4$), and a gas-cell (lamp) path-length of $l=1.5$\,m. We present the enhanced PNC optical rotation $2N\phipnc(\omega)$, multiplied by the transmission, which depends on the absorptivity of the specified transition through the atomic medium (Sec.\,\ref{sec:IV}, A).\\
\begin{figure}[h!]
\begin{center}
\includegraphics[angle=0, width=0.95\linewidth]{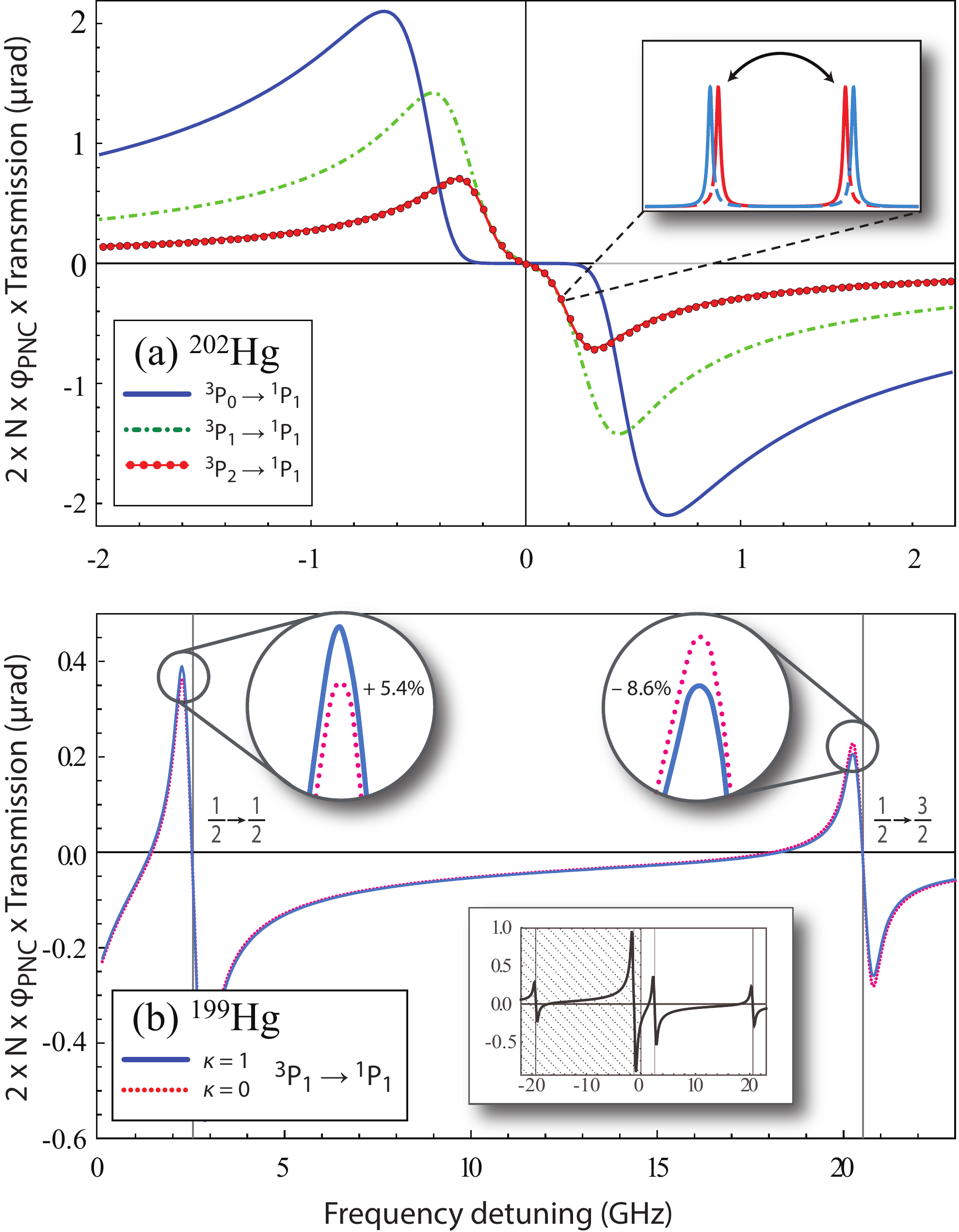}\\[-1.4ex]
\end{center}
\caption{\small{(color online). Theoretical simulations of the PNC optical rotation signal in Hg vs optical frequency for two cases. (a) Simulations for the three proposed transitions (transition wavelength $\lambda=609$, 682, 997\,nm) assuming a discharge lamp filled with isotopically pure $^{202}$Hg. For all the initial states, $^3P_0$, $^3P_1$ and $^3P_2$, we use densities $\rho=5\times10^{12}$\,cm$^{-3}$. In addition, identical Lorentz line-widths $\Gamma_{\rm L}=2\pi\times100$\,MHz for all transitions are used, while the Doppler line-widths for the 609, 682, and 997\,nm transitions are $\sim2\pi\times$\,183, 163, and 112\,MHz respectively. The (red) points in the $^3P_2\rightarrow{}^1P_1$ transition are separated by one FSR ($2\pi\times40$\,MHz), and the inset shows the reversal mechanism from allowing alternation between different polarization pairs yielding a net
polarization difference of $2\text{N}\phipnc$. (b) Isotopically pure odd-isotope $^{199}$Hg with $\rho=5\times10^{12}$\,cm$^{-3}$. The inset shows the full hyperfine structure of the transition. The effect of the nuclear anapole moment is presented, setting $\varkappa=1$ to yield visibly large signal differences. In each case we assume an isotopically pure filled discharge lamp filled, and density $\rho=5\times10^{12}$\,cm$^{-3}$ and Lorentz contributions of $\Gamma_L=2\pi\times100$\,MHz. See text for a detailed discussion.}} 
\label{fig:Hg}
\end{figure}
\indent {\bf Hg}: In Fig.\,\ref{fig:Hg} we present the theoretical PNC optical rotation simulations for the proposed transitions in Hg (using the values for $\mathcal{R}$ from Ref.\,\cite{DzuFlaXeHg} as presented in Table\,\ref{t:pnc}). In Fig.\,\ref{fig:Hg} (a), we assume equal densities $\rho=5\times10^{12}$\,cm$^{-3}$ of pure $^{202}$Hg for all the initial states of the proposed PNC transitions ($^3P_0$, $^3P_1$ and $^3P_2$), produced in a discharge lamp (or using an optical pumping scheme). The line-shape is a Voigt profile, with a Doppler contribution in the line-width of $\Gamma_{\rm D}\simeq2\pi\times$\,267, 238, and 163\,MHz for the 609, 682, and 997\,nm transitions respectively (see Eq.\,\ref{eq:Doppler} for $\sim320$\,K). The Lorentzian contribution for all three lines was taken to be $\Gamma_{\rm L}=2\pi\times100$\,MHz. This assumption is based on the fact that in a low-pressure discharge lamp ($<10$\,mTorr), the pressure broadening mechanisms are negligible compared to other homogeneous broadening mechanisms\,\cite{Lawler}. Therefore, the main contributions come from radiative processes. Lines originating from the $^3P_J$ states have Lorentz line-widths in the order of 20\,MHz, and in the order of 100\,MHz for lines originating from the $^1P_1$ state\,\cite{Lawler}. Assuming an effective path-length of $150\times10^4$\,cm, we get column densities that correspond to 12, 3 and 3 absorption lengths for the 609, 682, and 997\,nm transitions, respectively. \\
\indent In Fig.\,\ref{fig:Hg}\,(b) we examine the nuclear spin-dependent PNC effects for the 682\,nm transition in $^{199}$Hg (nuclear spin $I=1/2$). Using the values calculated by Dzuba and Flambaum in Ref.\,\cite{DzuFlaXeHg} for the PNC amplitudes between different hyperfine components, and by setting $\varkappa=1$, we see that the peak signals differ by about $+5.4$\% and $-8.6$\% resulting in total signal differences of up to $\sim14$\%. The actual value of $\varkappa$ can be estimated using Eq.\,\eqref{eq:varkappa} and Eq.\,\eqref{eq:anapole} to be $\sim0.1$ for the Hg nucleus. Therefore, achieving an experimental precision of at least 0.25\% is necessary to measure the NSD-PNC effects with a 6$\sigma$ precision, in the 682\,nm transition for $^{199}$Hg. Note that, similarly to the case of I\,\cite{PNCI}, the PNC signals for the two hyperfine groups, $F\,=\,1/2\rightarrow F^{\prime}\,=\,1/2$ and $F\,=\,1/2\rightarrow F^{\prime}\,=\,3/2$, deviate in opposite directions, a signature that serves as an important experimental check.\\
\begin{figure}[h!]
\begin{center}
\includegraphics[angle=0, width=0.95\linewidth]{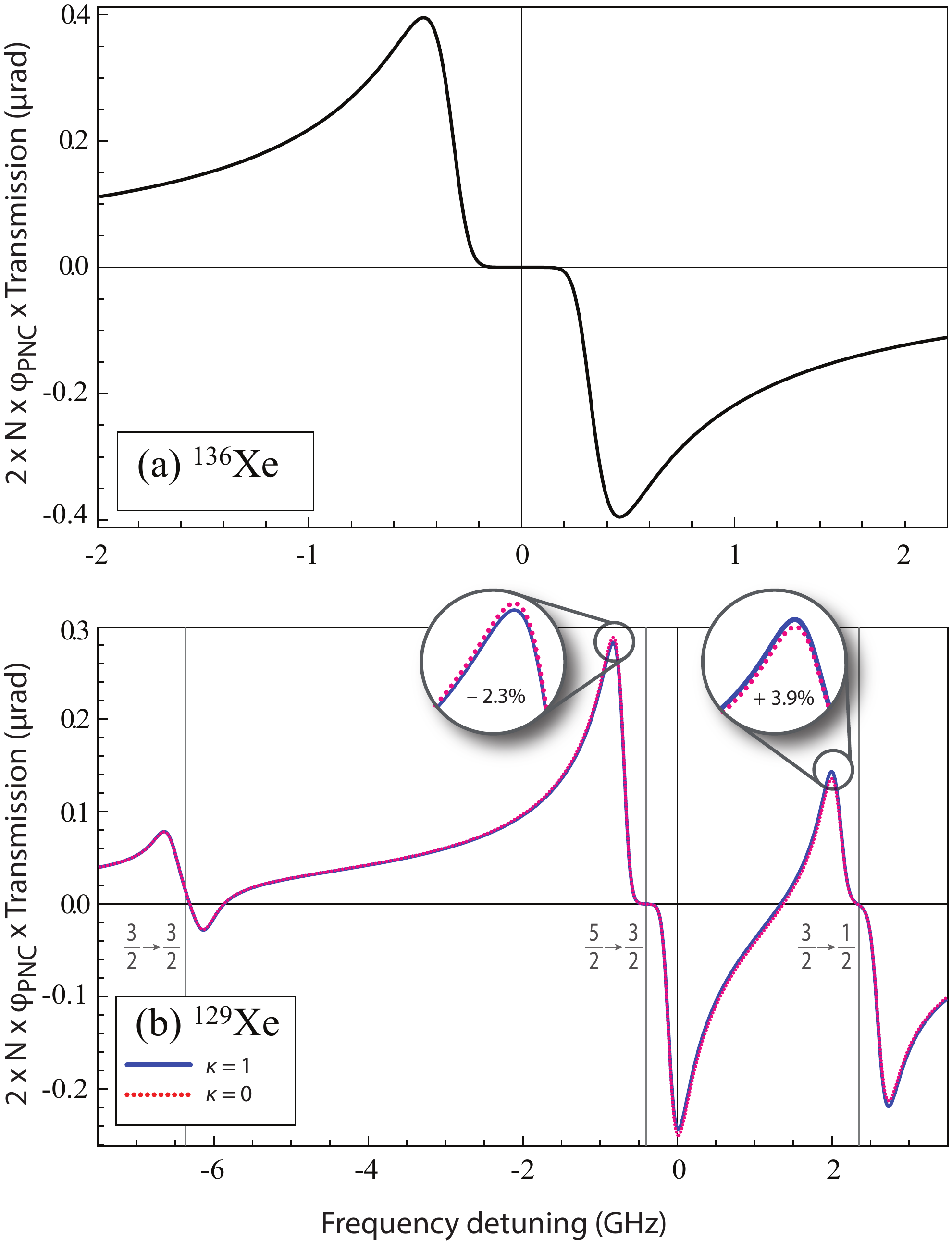}
\end{center}
\caption{\small{(color online). Theoretical prediction of the PNC optical rotation signal for metastable Xe vs optical frequency. (a) For a discharge lamp filled with isotopically pure metastable $^{136}$Xe, of density corresponding to 12 absorption lengths. (b) The (red) points in the $^3P_2\rightarrow{}^1P_1$ transition are separated by one FSR ($2\pi\times40$\,MHz). See text for a detailed discussion. All values taken from Ref.\,\cite{DzuFlaXeHg}. }}
\label{fig:Xe}
\end{figure}
\begin{figure}[h!]
\begin{center}
\includegraphics[angle=0, width=0.95\linewidth]{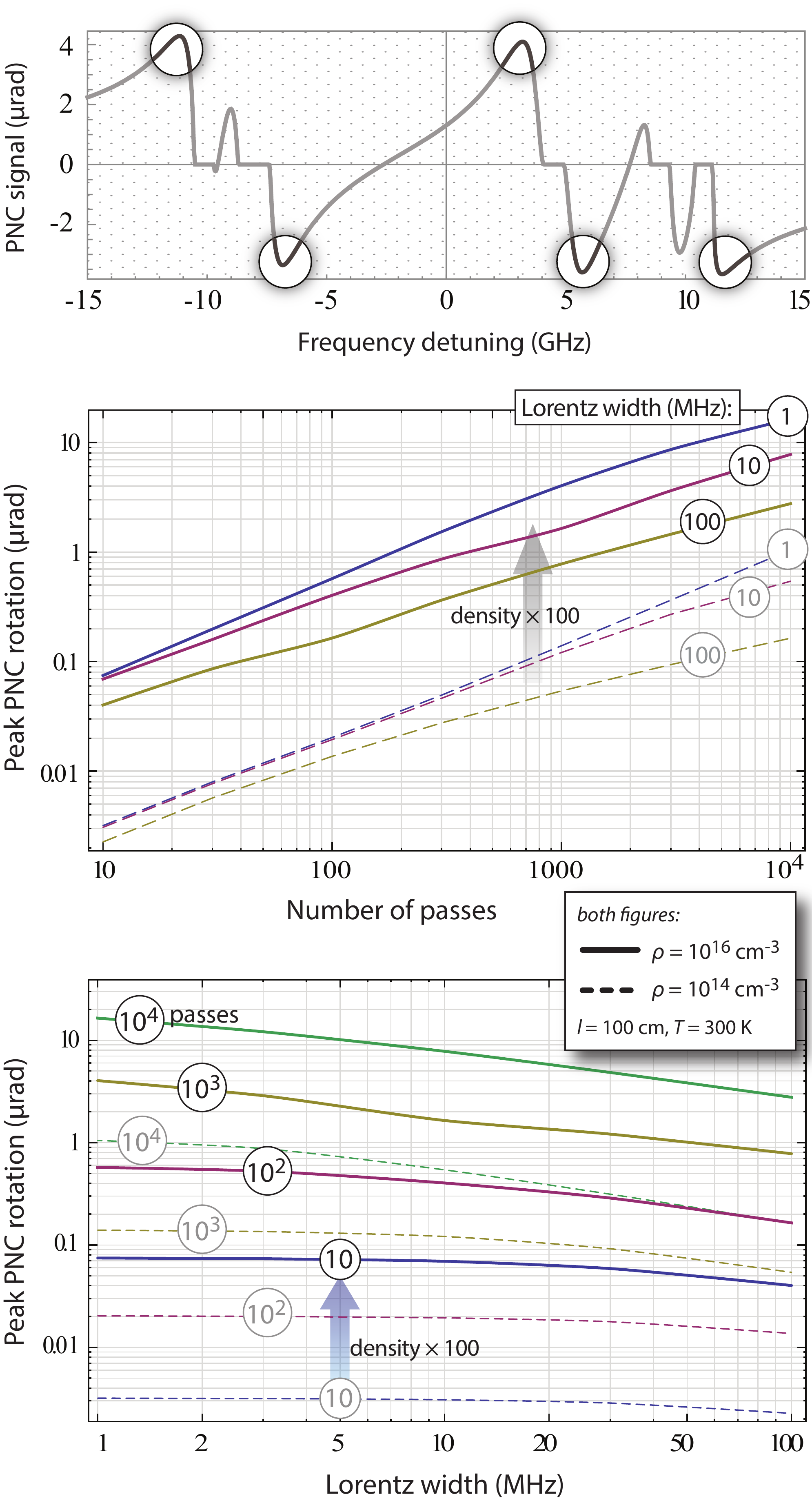}
\end{center}
\caption{\small{(color online) \emph{(top figure)} Theoretical prediction of the PNC optical rotation signal for the $^{2}$P$_{3/2}\rightarrow ^{2}$P$_{1/2}$ transition in $^{127}$I (we assume a column density of $\rho l=1.75\times10^{21}$\,cm$^{-2}$, and $\Gamma_{\rm L}=2\pi\times 3$\,MHz). \emph{(lower figures)} Calculations of the maximum (peak) PNC optical rotation angle are presented, as a function of the Lorentzian broadening of the line and the average number of passes $N$ (proportional to the finesse of the cavity). The simulations are performed for two different extreme-case densities, $\rho=10^{14}$ and $10^{16}$\,cm$^{-3}$, assuming constant interaction path-length and temperature ($\Gamma_{\rm D}=2\pi\times151$\,MHz). The non-smooth features in the simulations are the result of the fact that the peak rotation is not always associated with the same hyperfine component, but switches between hyperfine components (depicted with black circles in the top figure).}}
\label{fig:Iodine}
\end{figure}
\indent {\bf Xe}: In Fig.\,\ref{fig:Xe} the theoretical simulations for the expected PNC rotation signals for metastable Xe are presented. In the simulations presented in Fig.\,\ref{fig:Xe} (a) we assume densities of $\rho=1\times10^{12}$\,cm$^{-3}$ of $^{136}$Xe, which can be produced produced in a discharge lamp. The Doppler width is $\Gamma_{\rm D}\simeq2\pi\times192$\,MHz (300\,K) and the Lorentz width $\Gamma_{\rm L}\simeq2\pi\times60$\,MHz, based on preliminary measurements on a low-pressure He-Xe discharge lamp performed in our lab, and on measurements presented in Ref.\,\cite{Busshian}. Assuming $l_{\rm eff}=150\times10^4$\,cm, we calculate column densities that correspond to 12 absorption lengths at the center of the absorption. Fig.\,\ref{fig:Xe}\,(b) shows the PNC optical rotation signal for the case of pure $^{129}$Xe (with nuclear spin $I=1/2$) demonstrating a resolved hyperfine structure. Assuming the same density, Doppler and Lorentz width as in the simulations for the $^{136}$Xe, we obtain column densities that correspond to 7 absorption lengths (at maximum absorption). Similarly to Hg, we set $\varkappa=1$ to see the experimental sensitivity to NSD effects. Using the values from Ref.\,\cite{DzuFlaXeHg}, we see a total signal difference of up to $\sim$\,6.2\%. As the actual value of $\varkappa$ is expected again to be $\sim0.1$ (Xe has an odd-neutron nucleus), an experimental precision of about 0.1\% (6$\sigma$ precision) is required to measure the nuclear anapole moment in Xe.\\
\indent In addition, Hg and Xe have large distributions of stable isotopes ($\Delta N/N=8/120$ and 12/76 respectively). Ratios of atomic PNC measurements along an isotope chain of the same element, can exclude large errors associated with atomic-structure effects\,\cite{Dzuba1986} and are sensitive to variations in the neutron distribution\,\cite{Fortson1990, Brown2009}. \\
\indent {\bf $^{127}$I}: In Ref.\,\cite{PNCI}, investigations of the expected PNC optical rotation signal in the 1315\,nm transition in $^{127}$I were presented. Here we explore further the range of experimental conditions, for which a measurable PNC optical rotation signal is achievable. In Fig.\,\ref{fig:Iodine} we present the maximum (peak) PNC-optical rotation angle as a function of the Lorentzian broadening of the line and the average number of passes $N$ (proportional to the finesse of the cavity) for two different extreme-case densities, $\rho=10^{14}$ and $10^{16}$\,cm$^{-3}$ (the former is the minimum density needed to produce observable PNC signals and the latter is the largest that can be produced using the photodissociation method\,\cite{PNCI}). Note that the peak optical rotation is not always associated with the same hyperfine component, but switches between hyperfine components depending on the experimental conditions. This peak switching is responsible for the kinks present in the curves of Fig.\,\ref{fig:Iodine}. Finally, we propose the production of these densities from the photodissociation of $I_2$ with 532\,nm radiation (see relevant discussion in Ref.\,\cite{PNCI}).\\%
\indent Using the values presented in Fig.\,\ref{fig:Iodine}, we see that for densities of $\rho=10^{16}$\,cm$^{-3}$, a Lorentzian contribution of $\Gamma_{\rm L}=2\pi\times10$\,MHz, and 400 average number of passes, a peak $\phipnc^{\rm max}$ optical rotation angle of $\sim1\,\mu$rad is expected. Setting $\varkappa=1$, we observe NSD-PNC signal differences of about $\sim8.5$\%. Using the previously measured values of $\kappa$ for Cs\,\cite{Wieman} as the expected value for the anapole moment in iodine ($\varkappa(^{127}{\rm I})~ \simeq ~ -\varkappa(^{133}{\rm Cs})  \simeq  ~-0.38(6)$), we see that a measurement of about $\sim$0.5\% sensitivity, corresponding to a 5\,nrad detection sensitivity, is required to measure the NSD-PNC effects in $^{127}$I with a 6$\sigma$ precision (see also discussion in Ref.\,\cite{PNCI}). \\
%

\section{Conclusions}
\indent In this article we presented the fundamental elements of a cavity-enhanced polarimetric measurement of PNC optical rotation. The polarization eigenstates of a four-mirror bow-tie cavity supporting counter propagating beams were presented. We demonstrate how an absolute measurement of the PNC optical rotation is possible even in the presence of linear birefringence. The measurement procedure and the availability of robust subtraction procedures using two distinct signal reversals were also discussed. Furthermore, theoretical simulations for the expected PNC optical rotation signals, utilizing the cavity-enhanced optical rotation technique under experimentally feasible parameters, were presented. These suggest that, for the proposed systems and experimental conditions, measurements of odd-neutron and odd-proton NSD-PNC effects are experimentally feasible. In addition, all the proposed systems are suitable for PNC measurements along a chain of isotopes, particularly Xe that has the largest distribution of stable isotopes. Finally, we demonstrate that particularly for the case of $^{127}$I, large optical rotation signals are expected. We argue that the proposed experimental conditions, and the corresponding expected signal values and detection sensitivities for the proposed transition in iodine, compare favorably to those of successful PNC optical-rotation experiments\,\cite{Vetter,McPherson,Meekhof1}, suggesting that iodine is the most favorable candidate for future PNC optical rotation experiments, currently pursued in our laboratory.\\
\acknowledgements
The authors would like to thank Dr. Ren\`e Bussiahn for providing the discharge lamp, on which preliminary measurements were performed, and for helpful discussions, and V. A. Dzuba and V. V. Flambaum for supporting atomic structure calculations and discussions. LB thanks Annie Clark for fruitful discussions. The work was supported by the European Research Council (ERC) grant TRICEPS (GA No. 207542), by the National Strategic Reference Framework (NSRF) grant Heracleitus II (MIS 349309-PE1.30) co-financed by EU (European Social Fund) and Greek national funds.

\appendix
\section{Lineshapes}\label{app:LorDop}
\setcounter{subsubsection}{0}
In the absence of inhomogeneous broadening mechanisms and for frequency detunings much smaller than the resonance frequency, $\mid \w - \wo\mid\ll \wo$, the dispersive and absorptive parts of the lineshape function take the familiar Lorentzian form:
\begin{align} 
\label{eq:lorentzD}\L'(\w-\wo)  &= \frac{1}{\pi}\frac{\w - \wo}{\left(\w-\wo\right)^2+ \left(\G/2\right)^2}\\
\label{eq:lorentzA}\L''(\w-\wo) &=\frac{1}{\pi}\frac{\G/2}{\left(\w-\wo\right)^2+ \left(\G/2\right)^2}
\end{align}
%
%
\noindent In a thermal vapor the Doppler broadening of the transition due to the motion of the atoms can not be neglected. The natural way to include it would be to substitute the frequency variable, $\w$, by its Doppler shifted value, $\w-\mathbf{k} \cdot\boldsymbol\upsilon$, where $\mathbf{k}$ is the wavenumber and $\boldsymbol\upsilon$ the atomic velocity, and integrate the Lorentzians over a Maxwell velocity distribution, thus arriving at what is known as the \emph{Voigt} profile. However, this convolution of a Lorentzian with a Gaussian distribution is, for computational purposes, more conveniently expressed through the \emph{Faddeeva} function, $w(z)$, which is a scaled complementary error function of a complex variable, $z=x+iy$:
\begin{equation}\label{eq:faddeeva}
w(z)=e^{-z^2} {\rm Erfc}(-iz)=w'(x,y)+i\:w''(x,y)
\end{equation}
\noindent For an atom of mass $M$ and for a transition centered at $\wo$, the Doppler half-width at $^1\!/_e$ is:
\begin{equation}\label{eq:Doppler}
\Dop = \wo \sqrt{\frac{2 k_{\rm B}T}{M c^2}}
\end{equation}
\noindent and the absorptive and dispersive parts of the lineshape are related to the real ($w'$) and imaginary ($w''$) parts of the Faddeeva function, respectively, via:
\begin{align} 
\label{eq:voigtA}\L''(\w-\w_{\rm o})\! &\rightarrow \!\V''(\w-\wo)=\frac{w'(\frac{\w-\w_{\rm o}}{\Dop},\frac{\G/2}{\Dop})}{\sqrt{\pi}\:\Dop}\\
\label{eq:voigtD}\L'(\w-\w_{\rm o})\! &\rightarrow\! \V'(\w-\wo)=\frac{w''(\frac{\w-\w_{\rm o}}{\Dop},\frac{\G/2}{\Dop})}{\sqrt{\pi}\:\Dop}.
\end{align}
%

%
\section{Index of refraction}\label{app:n}
\setcounter{subsubsection}{0}
\subsubsection{$E2$ - Electric quadrupole interaction}
In our proposed transitions for Xe, Hg and I (with the exception of the $^3$P$_0^{\rm o}\rightarrow^1$P$_1^{\rm o}$ transition in Hg) selection rules allow for the existence of an electric quadrupole interaction which must be included. The electric quadrupole operator for the $q=\pm 1$ component of polarization is:
\begin{equation}\label{eq:quadop}
-\frac{q \w}{4 \sqrt{3}}Q^{(2)}_q,~~{\rm where}~~ Q^{(2)}_q=-2 e r^2\sqrt{\frac{4\pi}{2\!\times\! 2+1}}~~ Y^{(2)}_{q}.
\end{equation}
Inclusion of the $E2$ electric-quadrupole amplitude to the index of refraction (in addition to the inclusion of the $E1_{\rm PNC}$ dipole amplitude), is performed by the substitution in Eq. \eqref{eq:nmatel}:
\begin{equation}
\frac{M1^2}{3}\rightarrow \mid\!\JRME{q\, i\, d^{(1)}_q+\mu^{(1)}_q-\frac{q \w}{4 \sqrt{3}}Q^{(2)}_q}\!\mid^2.
\end{equation}
\indent Introducing the electric quadrupole to magnetic dipole ratio parameter, $\chi$:
\begin{equation}
\chi = \frac{\w}{4\sqrt{3}} \frac{\JRME{Q^{(2)}}}{\JRME{\mu^{(1)}}},
\end{equation}
and using (assuming that space is isotropic)
\begin{equation}\label{eq:isotropy}
\JRME{T^{(k)}_q} = \frac{1}{2k+1}\JRME{T^{(k)}},
\end{equation}
\noindent we arrive at (by use of Eq.\,\eqref{eq:FvsJRME}):
\begin{equation}\label{eq:nFQ}
n = 1 + n_{\rm o}\:\sum_{F,F'} C'_{FF'}\: \V_{FF'}(\w),
\end{equation}
\noindent the difference with Eq. \eqref{eq:nF} being the $C_{FF'}\rightarrow C'_{FF'}$ substitution with:
\begin{align}\label{eq:CFFQ}
C'_{FF'} &= \frac{\left(2 F+1\right)\left(2 F'+1\right)}{2 I+1} \nonumber\\ 
&\times \left(\sixj{J}{1}{J'}{F'}{I}{F}^2 + \frac{3 \chi^2}{5}\sixj{J}{2}{J'}{F'}{I}{F}^2\right).
\end{align}
\indent Note that no interference term between the electric quadrupole and PNC dipole interactions appears, as it cancels out when one explicitly performs the summation across the magnetic sublevels before reducing the matrix elements.\\
\subsubsection{Accounting for isotopes}
\indent In the case where the studied vapor comprises more than one isotopes, each with an abundance $b_i$, the index of refraction will just be the sum of the refractive indices for each isotope, $n_i$, weighted by their respective abundances:
\begin{equation}\label{eq:niso}
n = \sum_i b_i\: n_i.
\end{equation}
\noindent The central difference of the various $n_i$ is in the resonance frequency $\w_{FF'}\rightarrow\w_{FF',i}$. Each isotope has an \emph{isotope shifted} resonance frequency, stemming from the slight variations in the electron wavefunctions due to the different nuclear masses. This is the only difference for even isotopes which have no nuclear spin, hence $F\: (F')\rightarrow J\:(J')$ and $\w_{FF',i}\rightarrow \w_{JJ',i}$. For odd isotopes, the non-zero nuclear spin causes the appearance of hyperfine structure with different ground and excited state hyperfine constants for each odd isotope. It is then $\w_{FF',i} = \w_{JJ'} + \delta\w_{i} + \Delta\w_{F',i}^{\rm (HF)}-\Delta\w_{F,i}^{\rm (HF)}$, with:
\begin{widetext}
	\begin{equation}\label{eq:HF}
		\Delta\w^{\rm (HF)}_F=\frac{1}{2}A^{\rm (HF)} K + B^{\rm (HF)} \frac{\frac{3}{2}K(K+1)-2I(I+1)J(J+1)}{I(I-1)(2I-1)J(J-1)(2J-1)},~~{\rm with}~~K=F(F+1)-I(I+1)-J(J+1),
	\end{equation}
\end{widetext}
\noindent where $A^{\rm (HF)}$ and $B^{\rm (HF)}$ are the magnetic dipole and electric quadrupole hyperfine constants, respectively, and $\delta\w_{i} $ is the isotope shift. Other affected quantities are the Doppler width (which is proportional to $1/\sqrt{M_i}$ and is taken into account in the calculations) and the reduced matrix elements (where changes are generally very small).\\
\subsubsection{Absorption cross section}
From the exponent of Eq. \eqref{eq:Tr}, and using Eq. \eqref{eq:nFQ} and \eqref{eq:niso}, we get the expression for the absorption cross section:\\
\begin{equation}\label{eq:sigma}
\sigma(\w) = \sigma_{\rm o}\sum_i\sum_{F,F'} b_i\:C'_{FF'} \V''_{FF',i}(\w),
\end{equation}
where, as discussed in Appendix \ref{app:n}(b), $b_i$ is the abundance of isotope $i$, the $C'_{FF'}$ are the geometry factors of Eq. \eqref{eq:CFFQ} and $\V''_{FF'}$ is the absorptive part of the Voigt profile given in Eq. \eqref{eq:voigtA}. In the equation above, the \emph{integrated absorption cross section}, $\sigma_{\rm o}$, is: 
\begin{equation}\label{eq:sigmao}
\sigma_{\rm o} = \frac{\pi\mo\w_{JJ'}}{\hbar~c}\: \frac{1}{2 J+1}\: \frac{M1^2}{3}.
\end{equation}
\noindent Note that $\sigma_{\rm o}$ does not have units of area. Note also that, if the quadrupole interaction is neglected ($C'_{FF'}\rightarrow C_{FF'}$), then $\sum_i\sum_{F,F'} b_i\:C_{FF'} = 1$ and, since $\int_{0}^{\infty}\V''(\w) d\w = 1$, it is also $\int_{0}^{\infty}\sigma(\w) d\w = \sigma_{\rm o}$.\\

\end{document}